\documentclass[11pt]{article}
\usepackage{verbatim,amsmath,amssymb}
\usepackage{epsfig,float,color}
\usepackage{epstopdf}
\usepackage{geometry}
\usepackage{setspace}
\usepackage{wrapfig}
\usepackage{hyperref}
\usepackage[utf8]{inputenc}
\usepackage{graphicx}
\usepackage{flushend}
\usepackage{tikz,pgfplots}
\usepackage[linesnumbered,ruled]{algorithm2e}

\usepackage[font={footnotesize}]{caption}
\usepackage[font={footnotesize}]{subcaption}
\usepackage{footnote}
\usepackage{multicol}
\usepackage{multirow}
\usepackage{enumitem}   
\usepackage{soul}
\usepackage{framed}
\usepackage[titletoc,title]{appendix}
\usepackage{bm}
\usepackage{siunitx}
\usepackage{cite}
\usepackage[makeroom]{cancel} % this is to cancle out and put the numer: \cancelto{0}{x}--x is cancelled to zero
\usepackage{natbib}
\usepackage{hyperref}
\usepackage{bigints} % provides bigger interger signs

\newenvironment{rcases}
{\left.\begin{aligned}}
	{\end{aligned}\right\rbrace}

\definecolor{darkblue}{rgb}{0,0,1}
\hypersetup{pdftex=true, colorlinks=true, breaklinks=true, linkcolor=blue, menucolor=blue, citecolor=blue, urlcolor=blue}
\newcommand{\pd}[2]{\frac{\partial #1}{\partial #2}}

\usepackage{tikz}
\usetikzlibrary{positioning, arrows.meta}
\tikzset{%
	myarrow/.style = {-Stealth, shorten >=5pt}
}

\begin{document}
	
	\begin{center}
		\Large{\bf{Topology optimization of stiff structures under self-weight for given volume using a smooth Heaviside function}}\\
		
	\end{center}
	
	\begin{center}

		\large{P. Kumar $^{\star,\,\dagger,}$\footnote{pkumar@mae.iith.ac.in; prabhatkumar.rns@gmail.com}}
		\vspace{4mm}
		
		\small{{$\star$}\textit{Department of Mechanical and aerospace Engineering, Indian Institute of Technology Hyderabad, 502285, India}}
			\vspace{4mm}
			
		\small{\textit{$\dagger$ \textit{Department of Mechanical Engineering, Indian Institute of Science,
					Bengaluru, 560012, Karnataka, India}}}\\
			\vspace{4mm}
			
 Published\footnote{This pdf is the personal version of an article whose final publication is available at \href{https://link.springer.com/article/10.1007/s00158-022-03232-x}{Structural and Multidisciplinary Optimization}}\,\,\,in \textit{Structural and Multidisciplinary Optimization}, 
			\href{https://link.springer.com/article/10.1007/s00158-022-03232-x}{DOI:10.1007/s00158-022-03232-x} \\
			Submitted on 19~November 2021, Revised on 09~March 2022, Accepted on 14~March 2022

	\end{center}
	
	\vspace{1mm}
	\rule{\linewidth}{.15mm}
	%ABSTRACT..............................................................
	{\bf Abstract:}
	  This paper presents a density-based topology optimization approach to design structures under self-weight load. Such loads change their magnitude and/or location as the topology optimization advances and pose several unique challenges, e.g., non-monotonous behavior of compliance objective, parasitic effects of the low-stiffness elements, and unconstrained nature of the problems. The modified SIMP material scheme is employed with the three-field density representation technique (original, filtered, and projected design fields) to achieve optimized solutions~close~to~0-1. A novel mass density interpolation strategy is proposed using a smooth Heaviside function, which provides a continuous transition between solid and void states of elements and facilitates tuning of the non-monotonous behavior of the objective.  A constraint that implicitly imposes a lower bound on the permitted volume is conceptualized using  the maximum permitted mass and the current mass of the evolving design.  Sensitivities of the objective and self-weight are evaluated using the adjoint-variable method. Compliance of the domain is minimized to achieve the optimized designs using the Method of Moving Asymptotes. The Efficacy and robustness of the presented approach are demonstrated by designing various 2D and 3D structures involving self-weight. The proposed approach maintains the constrained nature of the optimization problems and provides smooth and rapid objective convergence. \\
	
	{\textbf {Keywords:} Topology optimization; Self-weight; Design-dependent loads; Heaviside projection function; Compliance minimization}

	\vspace{-4mm}
	\rule{\linewidth}{.15mm}
	%%%%%%%%%%%%%%%%%%%%%%%%%%%%%%%%%%%%%%%%%%%%%%%%%%%%%%%%%%%%%%%%%%%%%%%%%%%%%%%%%%%%%%%%%%%%%%%%%%%%%%%%%%%%%%%%%%%%%%%%%%%%%%%%%%%%%%%%%%%
	
\section{Introduction}
Topology optimization (TO) is gaining popularity constantly as a design tool to find optimized material distributions for a wide range of problems, including  single- and/or multi-physics concepts \citep{sigmund2013topology}. Problems involving their self-weight loads, \linebreak design-dependent forces  \citep{kumar2020topology}, provide various distinctive challenges \citep{bruyneel2005note}, e.g.,  (1)~non-monotonous characteristics of the compliance objective with respect to the design variables, (2)~tendency to lose constrained nature of the compliance optimization problems with  given volume constraints and (3)~the parasitic effects of low-stiffness elements. Note that parasitic effects are also observed in the eigenvalues maximization problem with the Solid Isotropic Material with Penalization (SIMP) \citep{pedersen2000maximization}. Civil engineering structures typically encounter self-weight loads and thus, their performances are directly associated with the location of optimized material distributions \citep{bruyneel2005note}.  In addition, consideration of the self-weight may be essential for designing large-scale structures. Herein, the motif is to present a density-based TO approach covering all the aforementioned challenges to optimize 2D and 3D structures subjected to self-weight. In a typical density-based TO approach, each finite element (FE) is assigned a design variable (material density) $x \in [0,\,1]$ that is assumed to be constant within the element.  $x= 1$ and $x = 0$ indicate the solid and void phases of the element, respectively.
\begin{figure}[h!]
	\begin{subfigure}[t]{0.50\textwidth}
	\centering
	\includegraphics[scale=1]{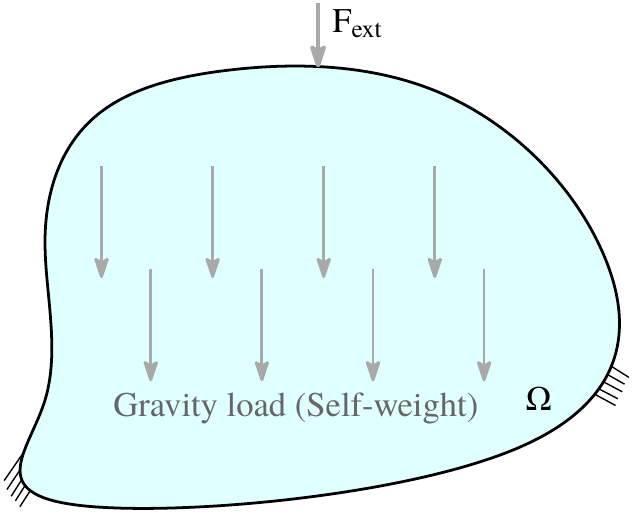}
	\caption{}
	\label{fig:sec1fig1a}
\end{subfigure}
~ \qquad
\begin{subfigure}[t]{0.5\textwidth}
	\centering
	\includegraphics[scale=1]{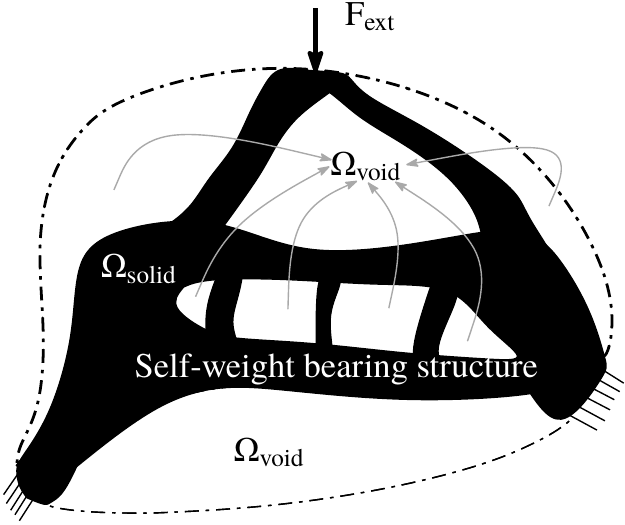}
	\caption{}
	\label{fig:sec1fig1b}
\end{subfigure}
	\caption{(\subref{fig:sec1fig1a}) Schematic diagram of a 2D design optimization problem with self-weight and an external load, F$_\text{ext}$. Presence of an external load may not always be necessary. (\subref{fig:sec1fig1b}) A representative solution to the design problem shown in (\subref{fig:sec1fig1a}). $\mathrm{\Omega},\,\mathrm{\Omega}_\mathrm{solid} (\bar{x}=1)\,\,\text{and}\, \mathrm{\Omega}_\mathrm{void} \,(\bar{x}=0)$ represent the actual design domain, the optimized material layout design (self-weight bearing structure) and void domain, respectively. $\bar{x}_e$ indicates the physical material density of element~$e$ in the parameterized setting.}\label{fig:sec1fig1}
\end{figure}

Design-dependent loads typically alter their location, magnitude and/or direction as TO progresses and therefore, their sensitivities  with respect to the design variables need to be considered within TO formulation \citep{kumar2020topology,kumar2020topology3Dpressure}. In such loading scenarios, the overall sensitivities of the compliance objective  with respect to the design variables no more remain always  negative (see Sec.~\ref{sec:Sensitivity analysis}) and hence, compliance loses its most cherished monotonic characteristics in a TO setting with self-weight \citep{bruyneel2005note}. In addition, when the effects of self-weight dominate, the optimization problem becomes unconstrained, and the corresponding optimized designs may be nonsensical in the mechanical viewpoint. Further, as per \cite{bruyneel2005note}, the parasitic effects are due to low-stiffness elements wherein the ratio between---gravity force to  design variables and stiffness to design variables tend to become unbounded.  The proposed approach offers a solution to the above-mentioned anomalies while providing optimized solutions close to 0-1 and subduing parasitic effects. A schematic diagram for a 2D structure\footnote{2D structure diagram is taken for simplicity.} experiencing self-weight (gravity load) and also, an external constant load, F$_\text{ext}$, is depicted in Fig.~\ref{fig:sec1fig1a}. Presence of an external load may not always essential though. A possible solution to the problem (Fig.~\ref{fig:sec1fig1a}) is displayed in Fig.~\ref{fig:sec1fig1b}. It can be noted that  location and magnitude of the  gravity load  change, which pose challenges in a TO formulation. Next, we review TO approaches presented for designing structures subjected to self-weight. 

In a structural optimization framework,  self-weight  was accounted first by  \cite{rozvany1977optimal}. Subsequently, several researchers have presented different TO approaches including self-weight. \cite{bruyneel2005note} identified challenges to include self-weight in  continuum-based TO settings. With a linear mass density interpolation, they found that when the design variables attain their lower bounds then the ratio between the mass density (gravity load or self-weight) and the material stiffness obtained by using the classical SIMP formulation tends to infinity. Consequently, displacements become unbounded and that in turn makes compliance unbounded. As per them, the optimizer tries to avoid  unboundedness of the displacement field by providing design variables in $(0<\bar{{x}}<1)$, i.e. gray elements. To subdue such unrealistic gray material, they  altered the SIMP formulation using a  threshold density value $\bar{x}_\text{th}$ and regarded material stiffness linear below $\bar{x}_\text{th}$, which introduces non-differentiability in the material model. This paper however shows that with the modified SIMP formulation \citep{sigmund2007morphology}, a smooth material (stiffness) interpolation scheme, in conjunction with the three-field density technique \citep{lazarov2016length}, one can circumvent parasitic effects.

\cite{ansola2006efficient} proposed a modified evolutionary structure optimization approach to design structures under self-weight. They presented a correction factor to compute the sensitivities of the objective and to enhance the convergence of the optimization. \cite{huang2011evolutionary}  used the bi-directional evolutionary structural optimization approach with the RAMP material model \citep{stolpe2001alternative} for designing structures subjected to self-weight. \cite{xu2013guide} proposed the guide-weight approach using the optimality criteria  method. They employed the SIMP and the RAMP material models to demonstrate their approach. \cite{chang2014gradient} presented a modified gradient projection method  to solve  problems involving density-dependent forces. \cite{holmberg2015worst} employed the non-linear semi-definite programming for the worst-case TO with self-weight. A closed B-splines-based approach was proposed by \cite{zhang2017cbs} to avoid parasitic effects of low-density regions for design-dependent loads.  \cite{felix2020topology} employed a power-law function for the material density interpolation to reduce the parasitic effects. \cite{fernandez2020simultaneous} presented a design optimization approach using the three-filed technique for simultaneously optimizing material, shape and topology. They used level-set functions to implicitly represent the boundaries and considered design-dependent pressure \citep{kumar2020topology} and self-weight loads. \cite{novotny2021topological} proposed a topological derivative-based TO approach  using a regularized formulation for imposing feasible volume constraints for structures under self-weight.

This paper presents an approach using the standard density-based TO with the modified SIMP scheme \citep{sigmund2007morphology} in conjunction with the three-field density representation technique (original, filtered and projected design fields, cf. \cite{lazarov2016length}) for the design problems involving self-weight. The parameter $\beta$ associated with the three-field technique is updated using a continuation scheme to achieve  optimized designs close to 0-1. A  novel mass density interpolation strategy is presented using a smooth Heaviside function. This interpolation scheme provides a continuous transition between the void and solid states of the element and controls the non-monotonous behavior of the objective. A constraint is formulated using  the maximum permitted self-weight (proportional to the given volume fraction) and the current self-weight of the evolving design. This constraint implicitly imposes a lower bound on the permitted volume. Consequently, the tendency to lose constrained nature of the problem is subdued. Load sensitivities are evaluated using the computationally cheap adjoint-variable method. Compliance is minimized to obtain optimized material layouts for various structures subjected to self-weight with different boundary specifications. We use the Method of Moving Asymptotes~\citep{svanberg1987method} to solve the formulated optimization problem. The approach can be readily extended for three-dimensional problems, which is demonstrated by solving two 3D numerical examples.

In summary, this paper offers the following new aspects:
\begin{itemize}
		\item A density-based topology optimization approach  using the three-field (original, filtered and projected) representation technique in conjunction with the modified SIMP scheme to design stiff structures subjected to self-weight. 
		\item Formulation of a novel mass density interpolation strategy using a smooth Heaviside function. The interpolation helps tune/control the non-monotonous behavior of the objective by using suitable mass density parameters (Sec.~\ref{sec:self-weight load modeling}). 
		\item Conceptualization of a new constraint within the optimization formulation that implicitly provides a lower bound on the permitted volume fraction for the given problem. When the effects of self-weight dominate, this constraint is necessary to retain the constrained nature of the problem (Sec.~\ref{sec:problem formulation}).
		\item Demonstration of the efficacy and robustness of the presented approach by designing various structures subjected to self-weight. Optimized solutions are close to 0-1 with smooth objective and constraints convergence, and thus parasitic effects are subdued (Sec.~\ref{sec: Numerical examples and discussions}).  
		\item Extension of the approach for 3D topology optimization design problems including self-weight (Sec.~\ref{sec: Numerical examples and discussions}). 
\end{itemize}

The layout of this paper is structured as follows. Section~\ref{sec:self-weight load modeling} describes the modeling of self-weight and the proposed mass density interpolation scheme.  Problem description including the topology optimization formulation and sensitivity analysis is presented in Section~\ref{sec:problem formulation}. Next, numerical examples for various 2D and 3D design problems involving self-weight are presented in Section~\ref{sec: Numerical examples and discussions}. Pertinent discussions and a study with different parameters are also presented. Section~\ref{sec:theedimension} demonstrates three-dimensional results. Lastly, conclusions are drawn in Section~\ref{sec:closure}.

\section{Self-weight modeling}\label{sec:self-weight load modeling}
Self-weight  $\mathbf{f}_\text{g}$ arises due to the gravitational acceleration $\mathbf{g}$, which acts vertically downward. In a continuum setting, $\mathbf{f}_\text{g}$  can be determined as 
\begin{equation}\label{eq:selfweightcontinuum}
	\mathbf{f}_\text{g} = \gamma V \mathbf{g} =  \gamma V \text{g}\, \mathbf{e},
\end{equation}
where $\gamma$ (\si{\kilogram \per \meter \cubed}) is  mass density of the material used, $V$ (\si{\meter\cubed}) represents volume of the domain, and $\mathbf{g} = \text{g}\mathbf{e} = -\SI{9.81}{\meter\per\square \second} \mathbf{e}$, where \text{g} $=-\SI{9.81}{\meter\per\square \second}$ and  $\mathbf{e}$, a unit vector,  directs in  positive $y-$ and $z-$directions for 2D and 3D settings respectively.

In a discrete setting,  the elemental self-weight $\mathbf{f}_\text{g}^e$  is evaluated as \citep{cook2007concepts}

\begin{equation}\label{eq:selfweightelement2D_3D}
	\mathbf{f}_\text{g}^e=\begin{cases}
		\bigints_{\Omega_e} \gamma_e \mathbf{N}^\top \begin{Bmatrix}
			0 \\ \text{g} \end{Bmatrix} dV = 		 \gamma_e\bigints_{\Omega_e}  \mathbf{N}^\top \begin{Bmatrix}
			0 \\ \text{g} \end{Bmatrix} dV, \, (\text{For 2D})\\
		\bigint_{\Omega_e} \gamma_e \mathbf{N}^\top \begin{Bmatrix}
			0 \\ 0\\ \text{g} \end{Bmatrix} dV = \gamma_e\bigint_{\Omega_e}  \mathbf{N}^\top \begin{Bmatrix}
			0 \\ 0\\ \text{g} \end{Bmatrix} dV,\, (\text{For 3D})
	\end{cases} \\
\end{equation}
where $\mathbf{N} = \left[N_1\mathbf{I},\,\cdots N_\text{Nno}\mathbf{I}\right]$. $\mathbf{I}$ is the identity matrix in $\mathcal{R}^n$, $n$ indicates dimension of the domain. $n =2$ and $n=3$ for 2D and 3D cases, respectively. $\gamma_e$ is the mass density of element $e$. Nno  is the total number of nodes per element employed to discretize the design domain. We use four-noded quadrilateral and eight-noded hexahedral  finite elements (FEs) ${\Omega_e}$ to describe the two- and three-dimensional design domains respectively, i.e, Nno~$=4$ (for 2D) and Nno~$=8$ (for 3D).

\begin{figure}[h!]
		\centering
		\begin{tikzpicture} 	
			\pgfplotsset{compat =1.9}
			\begin{axis}[
				width = 0.75\textwidth,
				xlabel= Physical variable ($\bar{{x}}$),
				ylabel= Mass density $\gamma_e\,(\si{\kilogram\per\meter\cubed})$ ,
				xtick={0,0.1,0.2,0.3,0.4,0.5,0.6,0.7,0.8,0.9,1.0},
				legend style={at={(0.95,0.65)},anchor=east}]
				\pgfplotstableread{MaterialDensity1.txt}\mydata;
				\addplot[smooth,black, mark = *,mark size=1pt,style={very thick}]
				table {\mydata};
				\addlegendentry{$\eta_\gamma = 0.01,\,\beta_\gamma = 10$}	
				\pgfplotstableread{MaterialDensity2.txt}\mydata;
				\addplot[smooth,red, mark size=1pt,style={very thick}]
				table {\mydata};
				\addlegendentry{$\eta_\gamma = 0.1,\,\beta_\gamma = 12$}
				\pgfplotstableread{MaterialDensity3.txt}\mydata;
				\addplot[smooth,blue, mark size=1pt,style={very thick}, dotted]
				table {\mydata};
				\addlegendentry{$\eta_\gamma = 0.2,\,\beta_\gamma = 8$}
				\pgfplotstableread{MaterialDensity4.txt}\mydata;
				\addplot[smooth,green, mark = square,mark size=1pt,style={very thick}]
				table {\mydata};
				\addlegendentry{$\eta_\gamma = 0.25,\,\beta_\gamma = 20$}	
			\end{axis}
		\end{tikzpicture}
	\caption{Mass density interpolation plot}
	\label{fig:Materialdensity}
\end{figure}
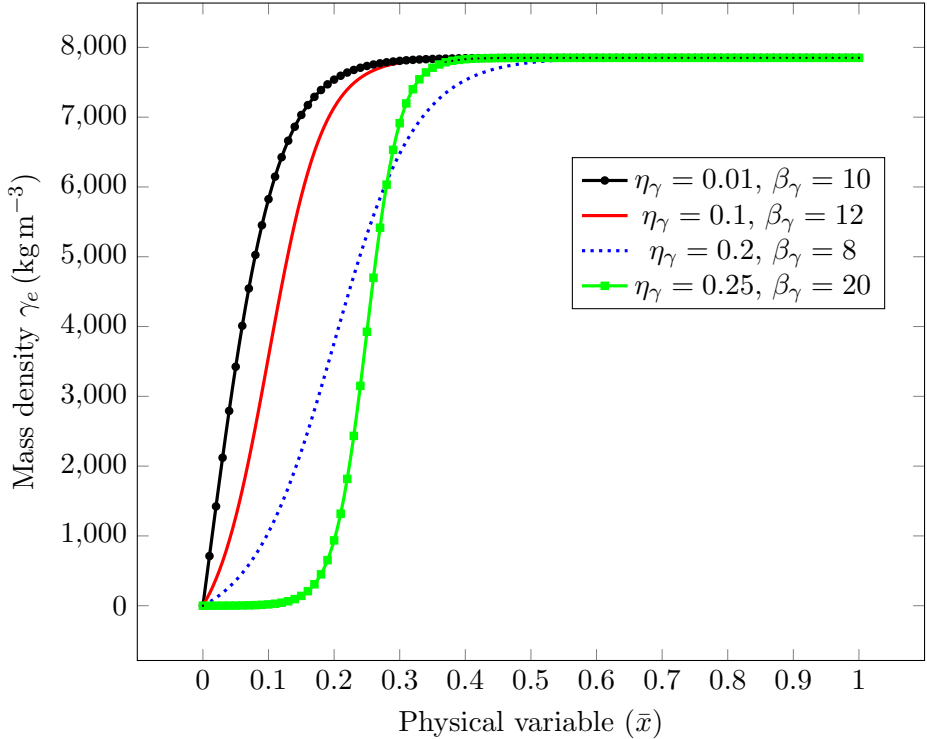

In a typical TO framework, each finite element displays solid ($\bar{x}_e = 1$) and void ( $\bar{x}_e = 0$) material phases; $\bar{x}_e$ denotes the physical design variable of element~$e$ (see Sec.~\ref{sec:problem formulation}). Herein, the mass density $\gamma_e$ of each FE  is related to  $\bar{x}_e$. Elements with $\bar{x}_e =0$ and $\bar{x}_e=1$  have $\gamma_e = \gamma_e^v$ (mass density of void FE) and $\gamma_e =\gamma_e^s = \gamma$ (mass density of solid FE) mass density respectively. A smooth Heaviside function is employed to evaluate the mass density of each FE using its both states that offers continuous transition between the phases of the element as TO progresses. 
In addition, the proposed interpolation scheme not only provides the load sensitivities readily (see Sec.~\ref{sec:Sensitivity analysis}), it offers a way to tune the non-monotonous behavior of the objective using the proper mass density parameters $\{\eta_\gamma,\,\beta_\gamma\}$ \eqref{eq:materialdensity} as described below. Mathematically, the mass density interpolation is written as  
\begin{equation}\label{eq:materialdensity}
	\begin{split}
\gamma_e &= \gamma_e^v + \left(\gamma_e^s -\gamma_e^v\right)H (\bar{x}_e,\,\eta_\gamma,\,\beta_\gamma),\\
&= \gamma_e^s \left(\chi + \left(1-\chi\right) H (\bar{x}_e,\,\eta_\gamma,\,\beta_\gamma) \right),
\end{split}
\end{equation}
where $\chi = \frac{\gamma_e^v}{\gamma_e^s} = 1e^{-9}$ is used; $\chi$ is termed the mass density contrast. The smooth Heaviside projection function $H (\bar{x}_e,\,\eta_\gamma,\,\beta_\gamma)$  is defined using the physical design variables as \citep{wang2011projection}
\begin{equation}\label{eq:Heavisidefunction}
	H (\bar{x}_e,\,\eta_\gamma,\,\beta_\gamma)  = \frac{\tanh(\beta_\gamma \eta_\gamma) + \tanh (\beta_\gamma(\bar{x}_e -\eta_\gamma))}{\tanh(\beta_\gamma \eta_\gamma) + \tanh (\beta_\gamma (1-\eta_\gamma))},
\end{equation}
where both $\eta_\gamma$ and $\beta_\gamma$ are adjustable parameters that provide the position of step and the slope. $\eta_\gamma$ and $\beta_\gamma$ are called the  mass density parameters herein. Fig.~\ref{fig:Materialdensity} depicts mass density interpolation function for different $\eta_\gamma$ and $\beta_\gamma$. For higher $\beta_\gamma$ more sharpness i.e. sharp transition from the void state to solid phase can be noted (Fig.~\ref{fig:Materialdensity}).

In view of shape functions of quadrilateral (2D) and hexahedral (3D) elements in association with the Gauss-quadrature points,  \eqref{eq:selfweightelement2D_3D} in light of \eqref{eq:materialdensity} gives  \citep{cook2007concepts}

\begin{equation}\label{eq:selfweightelementfinal1}
	\mathbf{f}_\text{g}^e  = \gamma_e^s \left(\chi + \left(1-\chi\right) H (\bar{x}_e,\,\eta_\gamma,\,\beta_\gamma) \right)L_g V_e,
\end{equation}
where $L_g = \left[0,\,\frac{\text{g}}{4},\,0,\,\frac{\text{g}}{4},\,0,\,\frac{\text{g}}{4},\,0,\,\frac{\text{g}}{4}\right]^\top$ and \begin{equation*}
	\begin{split}
	L_g=\left[0,\,0,\,\frac{\text{g}}{8},\,0,\,0,\,\frac{\text{g}}{8},\,0,\,0,\,\frac{\text{g}}{8},\,0,\,0,\,\frac{\text{g}}{8},\,0,\,0,\,\frac{\text{g}}{8},\,0,\,0,\,\frac{\text{g}}{8},\,\right.\\ \left. 0,\,0,\,\frac{\text{g}}{8},\,0,\,0,\,\frac{\text{g}}{8}\right]^\top \end{split} 
\end{equation*}for 2D and 3D settings respectively, and $V_e = \frac{V}{Nel} = \frac{L_x L_y t}{\texttt{Nel}}$ (for 2D), $V_e = \frac{V}{Nel} = \frac{L_x L_y L_z}{\texttt{Nel}}$ (for 3D). $L_x$, $L_y$, $L_z$ and $t$ represent the length, width, height and thickness of the design domain. The derivatives of gravity force $\mathbf{f}_\text{g}^e$ with respect to  $\bar{x}_e$ can be readily evaluated as
\begin{equation}\label{eq:loadsenstivities}
	\pd{\mathbf{f}_\text{g}^e }{\bar{{x}}_e} = \gamma_e^s \left(\chi + \left(1-\chi\right) \pd{H(\bar{x}_e,\,\eta_\gamma,\,\beta_\gamma)}{\bar{{x}}_e} \right)L_g V_e,
\end{equation}
where
\begin{equation*}\label{eq:smoothheavisidederivative}
	\pd{H(\bar{x}_e,\,\eta_\gamma,\,\beta_\gamma)}{\bar{x}_e} = \beta_\gamma \frac{1 - \tanh\left(\beta_\gamma(\bar{x}_e -\eta_\gamma)\right)^2}{\tanh(\beta_\gamma \eta_\gamma) + \tanh (\beta_\gamma (1-\eta_\gamma))}.
\end{equation*}
Finally, the elemental forces are assembled to determine the total (global) gravity load $\mathbf{F}_\text{g}$ experienced by the  design in a discrete setting and thus, compliance of the domain and
  $\pd{\mathbf{F}_\text{g}}{\bar{\mathbf{x}}}$ are evaluated. 
  
  \begin{figure}
  	\centering
  	\begin{tikzpicture} 	
  		\pgfplotsset{compat =1.9}
  		\begin{axis}[
  			width = 0.75\textwidth,
  			xlabel= Physical variable ($\bar{{x}}$),
  			ylabel=  $-\pd{\text{f}_\text{g}^e}{\bar{x}_e}$ ,
  			xtick={0,0.1,0.2,0.3,0.4,0.5,0.6,0.7,0.8,0.9,1.0},
  			legend style={at={(.95,0.65)},anchor=east}]
  			\pgfplotstableread{DerivativeMaterialDensity1.txt}\mydata;
  			\addplot[smooth,black, mark = *,mark size=1pt,style={very thick}]
  			table {\mydata};
  			\addlegendentry{$\eta_\gamma = 0.01,\,\beta_\gamma = 10$}	
  			\pgfplotstableread{DerivativeMaterialDensity2.txt}\mydata;
  			\addplot[smooth,red, mark size=1pt,style={very thick}]
  			table {\mydata};
  			\addlegendentry{$\eta_\gamma = 0.1,\,\beta_\gamma = 12$}
  			\pgfplotstableread{DerivativeMaterialDensity3.txt}\mydata;
  			\addplot[smooth,blue, mark size=1pt,style={very thick}, dotted]
  			table {\mydata};
  			\addlegendentry{$\eta_\gamma = 0.2,\,\beta_\gamma = 8$}
  			\pgfplotstableread{DerivativeMaterialDensity4.txt}\mydata;
  			\addplot[smooth,green, mark = square,mark size=1pt,style={very thick}]
  			table {\mydata};
  			\addlegendentry{$\eta_\gamma = 0.25,\,\beta_\gamma = 20$}	
  		\end{axis}
  	\end{tikzpicture}
  	\caption{For a 2D case with $L_x = L_y = \SI{1}{\meter}$ and $t = \SI{0.01}{\meter}$, different plots of $\pd{\text{f}_\text{g}^e}{\bar{x}_e}$ with respect to $\bar{{x}}$, where $\text{f}_\text{g}^e$ is the $y-$component of  $\mathbf{f}_\text{g}^e$. $\eta_\gamma$ provides location of the peak of the curve on $\bar{x}$-axis, and $\beta_\gamma$ controls the sharpness of peak of the curve at $\eta_\gamma$. A high $\beta_\gamma$ gives a relatively sharp peak at its $\eta_\gamma$, which can jeopardise the optimization process. Sec.~4.2.2 provides a recommendation for $\{\eta_\gamma,\,\beta_\gamma\}$ for the given design problem.}
  	\label{fig:self-weightderivative}
  \end{figure}

As per \eqref{eq:loadsenstivities}, $\frac{\partial \mathbf{f}_\text{g}^e}{\partial \bar{x}_e}\propto \gamma_e^s \left(\chi + \left(1-\chi\right) \pd{H(\bar{x}_e,\,\eta_\gamma,\,\beta_\gamma)}{\bar{{x}}_e} \right)$, i.e., $\frac{\partial \mathbf{f}_\text{g}^e}{\partial \bar{x}_e}$ is a function of physical design variable $\bar{x}_e$ at desirable $\{\eta_\gamma,\,\beta_\gamma\}$ and thus, alters as TO progresses. This is expected to help the exploratory characteristics of the TO process \citep{kumar2020topology}. Fig.~\ref{fig:self-weightderivative} depicts different plots of $\pd{\text{f}_\text{g}^e}{\bar{x}_e}$ with respect to $\bar{x}$, where $\text{f}_\text{g}^e$ is the $y-$component of  $\mathbf{f}_\text{g}^e$, at the same sets of  $\{\eta_\gamma,\,\beta_\gamma\}$ those are used in Fig.~\ref{fig:Materialdensity}. $\eta_\gamma$ determines  the location of the peak of the derivative curve on $\bar{x}$-axis whereas, $\beta_\gamma$ controls the sharpness of the curve  at $\eta_\gamma$ (Fig.~\ref{fig:self-weightderivative}).  Elements with design variables lower than $\eta_\gamma$ have mass density close to $\gamma_v$ (Fig.~\ref{fig:Materialdensity}).  A higher $\beta_\gamma$ can give a relatively sharper peak at  $\eta_\gamma$ and thus, it can jeopardize the TO process. Therefore, for a given design problem, $\{\eta_\gamma,\,\beta_\gamma\}$ is selected so that used initial guess for the design vector, typically taken equal to the given volume fraction, remains at a proper distance  from $\eta_\gamma$ towards the right side on $\bar{{x}}$-axis wherein the derivatives of $\pd{\text{f}_\text{g}^e}{\bar{x}_e}$ are relatively much lower (Fig.~\ref{fig:self-weightderivative}). In this way, the non-monotonous behavior of the objective can be tuned for the given problem. Consequently, the optimization process becomes smooth and gives mechanical sensible designs (see Sec.~\ref{sec: Numerical examples and discussions}). We provide a recommendation to decide  $\{\eta_\gamma,\,\beta_\gamma\}$ for a given design problem in Sec.~\ref{sec:Massdensityparameters} based on the numerical experiments performed therein. Next,  the  problem formulation is presented.

\section{Problem formulation}\label{sec:problem formulation}

The density-based topology optimization  in conjunction  with the three-field density formulation \citep{lazarov2016length} is used in the presented approach. For the Young's modulus interpolation, the modified SIMP formulation \citep{sigmund2007morphology}  is employed that relates modulus of elasticity $E_e$ of element~$e$ to its physical design variable $\bar{x}_e$ using the power law as
\begin{equation}\label{eq:SIMP formulation}
	E_e(\bar{x}_e) = E_e^v + \left(E_e^s- E_e^v\right) (\bar{x})^p,
\end{equation}
where $E_e^v$ and $E_e^s$ are Young's moduli of the void and the solid phases of element $e$, respectively. The material contrast, i.e. $\frac{E_e^v}{E_e^s} = \SI{1e-6}{}$ is set and $p$, the SIMP penalty parameter, is set to 3  that guides TO  convergence towards \textquoteleft0-1' designs.

In the three-field ($\mathbf{x},\,\tilde{\mathbf{x}},\,\bar{\mathbf{x}}$) density representation technique \citep{lazarov2016length}, $\mathbf{x}$, $\tilde{\mathbf{x}}$ and $\bar{\mathbf{x}}$ denote vectors containing the original  design variables $x_e$, filtered design variables $\tilde{x}_e$ and physical design variables $\bar{x}_e$ respectively.  The chain of transformation between these variables can be denoted via ${x}_e \to\tilde{x}_e \to \bar{x}_e$ \citep{lazarov2016length}. 

The filtered variables $\tilde{x}_e$, determined using a mesh-independent density filtering scheme \citep{bruns2001topology}, is given as
\begin{equation}\label{eq:filteredvariables}
	\tilde{x}_e = \frac{\sum_{i=1}^{nne} v_i x_i w(\mathbf{z}_{e,i})}{\sum_{i=1}^{nne}v_iw(\mathbf{z}_{e,i})},
\end{equation}
where $nne$ indicates the total number of neighboring elements of element $e$, $v_i$ is the volume of the element~$i$.  $w(\mathbf{x}_i)$, the weight function, is determined using the Euclidean distance between the centroids $\mathbf{z}_e$ and $\mathbf{z}_i$ of elements $e$ and $i$ as
\begin{equation}\label{eq:weightfunction}
	w(\mathbf{z}_{e,i}) = \max \left(0,\, 1-\frac{||\mathbf{z}_e-\mathbf{z}_i||}{r_\text{fill}}\right),
\end{equation}
where $r_\text{fill}$ is the employed filter radius. One can write \eqref{eq:filteredvariables} in the matrix form as
	\begin{equation}
	\tilde{\mathbf{x}} = \mathbf{P\,x},\,\,\text{where,}\,\mathbf{P}_{(i,j)} = \begin{cases}
	\frac{v_j w(\mathbf{z}_{i,j})}{\sum_{k=1}^{nne} v_k w(\mathbf{z}_{i,k})},\,\, j\in \mathcal{N}_{i,j}\\
	0, \quad \text{otherwise }
	\end{cases}
	\end{equation}
	where $\mathcal{N}_{i,j}$ represents the set of neighboring elements for element $i$ within the given filter radius $r_\text{fill}$. We compute filter matrix $\mathbf{P}$ once in the beginning of the algorithm, store it as a sparse matrix and use it within the optimization loop to evaluate the filtered design vector $\tilde{\mathbf{x}}$.
The derivatives of $\tilde{x}_e$ with respect to $x_i$ is calculated as
\begin{equation}\label{eq:filteredderivative}
	\frac{\partial \tilde{x}_e}{\partial x_i} =  \frac{v_i  w(\mathbf{z}_{e,i})}{\sum_{j=1}^{nne}v_jw(\mathbf{z}_{e,j})},\,\text{ i.e.,}\,\, 	\frac{\partial \tilde{\mathbf{x}}}{\partial \mathbf{x}} = \mathbf{P}.
\end{equation}

The physical design variable $\bar{x}_e$, determined using corresponding filtered  variable $\tilde{x}_e$ and a smooth Heaviside function which is analogous to \eqref{eq:Heavisidefunction}, is given as
\begin{equation}\label{eq:physicaldensity}
	\begin{split}
		\bar{x}_e& = H (\tilde{x}_e,\,\beta,\,\eta)|_{\eta = 0.5}
		= \frac{\tanh (\frac{\beta}{2}) + \tanh (\beta(\tilde{x}_e-\frac{1}{2}))}{2\tanh(\frac{\beta}{2})}
	\end{split}
\end{equation}
where $\beta \in [1,\,\infty)$ controls the sharpness of the Heaviside function $H (\tilde{x}_e,\,\beta,\,\eta)$. Typically $\beta$ is increased in a continuation fashion  from its initial value $\beta_i =1$ to maximum value $\beta_\text{max}$ to achieve the solution close to 0-1 \citep{wang2011projection}. In turn,  parasitic effects of the low-density element can be suppressed. Herein, $\beta_\text{max} = 256$ is set, and  is doubled after every 25 optimization iterations. With $\eta = 0$ and $\eta =1$, one achieves the minimum length scale on the void and solid phase respectively \citep{wang2011projection}, and also, attains the Heaviside approximation given by \cite {guest2004achieving} and \cite{sigmund2007morphology} respectively. One determines the derivative of $\bar{x}_e$ with respect to $\tilde{x}_e$ as
\begin{equation}\label{eq:heavisidederivative}
	\frac{\partial \bar{x}_e}{\partial \tilde{x}_e} = \beta \frac{1 - \tanh(\beta(\tilde{x}_e -\frac{1}{2}))^2}{2\tanh(\frac{\beta}{2})}.
\end{equation}
Using the chain rule, the derivative of a function~$f$ with respect to the actual design variable can be determined in view of \eqref{eq:filteredderivative} and \eqref{eq:heavisidederivative} as
\begin{equation}\label{eq:chainrule}
	\frac{\partial f}{\partial x_i} = \sum_{e = 1}^{nne}\frac{\partial f}{\partial \bar{x}_e}\frac{\partial \bar{x}_e}{\partial \tilde{x}_e}\frac{\partial \tilde{x}_e}{\partial {x}_i},
\end{equation}
wherein $\frac{\partial f}{\partial \bar{x}_e}$ can be determined for the given objective function (see~Sec.~\ref{sec:Sensitivity analysis}).

\subsection{Topology optimization formulation}\label{sec:Topology optimization formulation}
The compliance (strain energy) of the structures subject to self-weight and/or external constant load is minimized to obtain the optimized topologies. The optimization problem is formulated as
\begin{equation}\label{eq:Optimizationequation}
	\begin{rcases}
		\begin{split}
			&\underset{\bar{\mathbf{x}}(\tilde{\mathbf{x}}(\mathbf{x}))}{\text{min}}\,\,f_0=2SE = {\left(\mathbf{F}_\text{g}(\bar{\mathbf{x}}) + \kappa\mathbf{F}_\text{ext}\right)}^\top \mathbf{u}(\bar{\mathbf{x}})\,\,\,\, \,\, \\
			&\text{Subject to: } \\
			& \bm{\lambda}: \qquad\,\mathbf{K}(\bar{\mathbf{x}})\mathbf{u}(\bar{\mathbf{x}}) = \mathbf{F}_\text{g}(\bar{\mathbf{x}}) + \kappa\mathbf{F}_\text{ext}\\
			& {\Lambda_1}: \qquad\,\text{g}_1\equiv V(\bar{\mathbf{x}})\le {V^*}\\
			&{\Lambda_2}: \qquad\,\text{g}_2\equiv   m_{\text{max}} \le \sum_{e=1}^{\texttt{Nel}}{m}_{e}\\
			& \quad \,\,\, \qquad\, \qquad\mathbf{0}\le\bar{\mathbf{x}}\le \mathbf{1}\\ 
			& \text{Data:} \quad V^*,\, \mathbf{F}_\text{ext},\, \gamma_s,\,g,\,L_x,\,L_y,\,L_z,\,t,\,E_e^s,\,\kappa
		\end{split}
	\end{rcases},
\end{equation}
where $f_0$ represents the objective function, i.e., compliance of the structure, $\mathbf{F}_\text{g}$ and $\mathbf{F}_\text{ext}$ are the global force vectors arise due to the self-weight and external loads respectively, and $SE$ indicates the strain energy. $\kappa$ is a user defined scalar quantity. $\mathbf{K}$ and $\mathbf{u}$ are the global stiffness matrix and the displacement vector respectively. $\text{g}_1$ and $\text{g}_2$ are constraints. $V(\mathbf{x})$ and $V^*$ are the design volume and permitted volume respectively. $V^* = V^*_f\times \texttt{Nel}$, where $V^*_f$ is the permitted volume fraction. $\bm{\lambda}$ (vector), $\Lambda_1$ (scalar) and $\Lambda_2$ (scalar) are the Lagrange multipliers corresponding to the state equations, g$_1$ and g$_2$ respectively.

Herein, constraint g$_2$ is applied such that a lower bound on the volume fraction can implicitly be realized. It is formulated using the maximum permitted mass determined using the given volume fraction, $m_\text{max} =V\gamma_sV^*_f$, and the intermediate mass of the evolving design, wherein $m_e = V_e \gamma_e$  indicates mass of element e.  Sec.~\ref{sec:qualify g2} and Sec.~\ref{sec:MBBbeamdesign} substantiate the requirement of this constraint  for a given volume when the effects of self-weigh dominate via numerical examples.  This constraint helps retain the constrained nature of the problem. For the finite element analysis, the small  elastic deformation  is assumed herein.

\subsection{Sensitivity analysis}\label{sec:Sensitivity analysis}
A gradient-based optimization technique, the Method of Moving Asymptotes (MMA, cf. \cite{svanberg1987method}), is employed  to solve the optimization problem mentioned in~\eqref{eq:Optimizationequation}. Thus, derivatives of the objective and constraints with respect to the design variables are required. The derivatives of objective are determined via the adjoint-variable approach. For that, the augmented response $\mathcal{L}$ is defined using the objective and the state equations as
\begin{equation}\label{eq:augmentedeqaution}
	\mathcal{L} = f_0(\mathbf{u}(\bar{\mathbf{x}})) + \bm{\lambda}^\top \left(\mathbf{K}(\mathbf{x})\mathbf{u}(\mathbf{x}) - \mathbf{F}_\text{g}(\mathbf{x}) - \kappa\mathbf{F}_\text{ext} \right),
\end{equation} 
where $\bm{\lambda}$ is the Lagrange multiplier vector. We henceforth for brevity omit the arguments from the vector and the matrix quantities. Differentiation of \eqref{eq:augmentedeqaution} with respect to $\bar{\mathbf{x}}$ yields
\begin{equation*}
	\begin{split}
		\frac{\text{d}{\mathcal{L}}}{\text{d}\bar{\mathbf{x}}} &= \pd{f_0}{\bar{\mathbf{x}}}+ \pd{f_0}{\mathbf{u}}\pd{\mathbf{u}}{\bar{\mathbf{x}}} + \bm{\lambda}^\top\left(\pd{\mathbf{K}}{\bar{\mathbf{x}}}\mathbf{u}+\mathbf{K}\pd{\mathbf{u}}{\bar{\mathbf{x}}} -\pd{\mathbf{F}_\text{g}}{\bar{\mathbf{x}}}\right)\\
		&= \pd{f_0}{\bar{\mathbf{x}}} + \underbrace{\left(\pd{f_0}{\mathbf{u}} + \bm{\lambda}^\top\mathbf{K}\right)}_{\Theta} \pd{\mathbf{u}}{\bar{\mathbf{x}}}+ \bm{\lambda}^\top\left(\pd{\mathbf{K}}{\bar{\mathbf{x}}}\mathbf{u} -\pd{\mathbf{F}_\text{g}}{\bar{\mathbf{x}}}\right).
	\end{split}
\end{equation*}
$\bm{\lambda}$ is selected such that $\Theta = 0$\footnote{This is the adjoint equation corresponding to the state equation  \eqref{eq:Optimizationequation}.} that yields, $\bm{\lambda} = -2\mathbf{u}$ and thus, with $f_0 = 2SE$ and at the equilibrium state, one writes
\begin{equation}\label{eq:objectivesensitivity}
	\begin{split}
		\frac{\text{d}f_0}{\text{d}\bar{\mathbf{x}}}
		&= \pd{f_0}{\bar{\mathbf{x}}} - 2\mathbf{u}^\top\left(\pd{\mathbf{K}}{\bar{\mathbf{x}}}\mathbf{u} -\pd{\mathbf{F}_\text{g}}{\bar{\mathbf{x}}}\right)\\
		&= -\mathbf{u}^\top \pd{\mathbf{K}}{\bar{\mathbf{x}}}\mathbf{u} + \underbrace{2\mathbf{u}^\top\pd{\mathbf{F}_\text{g}}{\bar{\mathbf{x}}}}_{\text{Self-weight sensivities}}.
	\end{split}
\end{equation}

In \eqref{eq:objectivesensitivity}, self-weight sensitivities, $2\mathbf{u}^\top\pd{\mathbf{F}_\text{g}}{\bar{\mathbf{x}}}$ appear and they are evaluated using \eqref{eq:loadsenstivities}. Next, the chain rule \eqref{eq:chainrule} is employed  to evaluate the objective sensitivities with respect to the design vector, i.e., $\frac{\text{d}f_0}{\text{d}\mathbf{x}}$. Finding sensitivity of constraint g$_1$ is straightforward \citep{sigmund2007morphology}, whereas that of constraint g$_2$ can be evaluated using~\eqref{eq:loadsenstivities}. Moreover, it can be noted that  compliance objective sensitivities can  either be negative or positive \eqref{eq:objectivesensitivity}, i.e., they alter their sign as per the different design variables. This shows the non-monotonous behavior of compliance when self-weight is considered. This behavior is controlled herein by the proposed mass density interpolation scheme. The selected (Sec.~\ref{sec:Massdensityparameters}) mass density interpolation parameters $\{\eta_\gamma,\,\beta_\gamma\}$  controls $\pd{\mathbf{F}_\text{g}}{\bar{\mathbf{x}}}$ (Fig.~\ref{fig:self-weightderivative}) of $2\mathbf{u}^\top\pd{\mathbf{F}_\text{g}}{\bar{\mathbf{x}}}$, and thus, the non-monotonous behavior. Note that when self-weight effects dominate, $2\mathbf{u}^\top\pd{\mathbf{F}_\text{g}}{\bar{\mathbf{x}}}$ always gives positive quantities. Sec.~\ref{sec:Massdensityparameters} provides a recommendation for choosing $\{\eta_\gamma,\,\beta_\gamma\}$ for the given optimization problem based on the numerical examples solved therein. Next, we solve various structure problems subjected to self-weight and provide pertinent  discussions.

\section{Numerical Examples and Discussions} \label{sec: Numerical examples and discussions}
\begin{figure}
	\begin{subfigure}[t]{0.45\textwidth}
		\centering
		\includegraphics[scale=1.25]{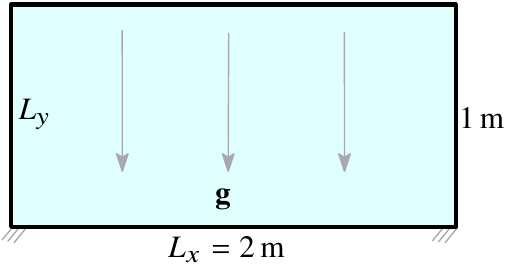}
		\caption{}
		\label{fig:problem 1}
	\end{subfigure}
	\quad
	\begin{subfigure}[t]{0.45\textwidth}
		\centering
		\includegraphics[scale=1.5]{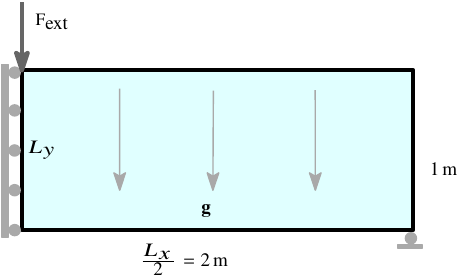}
		\caption{}
		\label{fig:problem 2}
	\end{subfigure}
	\caption{Design domains (\subref{fig:problem 1}) Arch geometry with dimension $L_x\times L_y = \SI{2}{\meter}\times \SI{1}{\meter}$  (\subref{fig:problem 2}) A symmetric MBB beam design with dimension $\frac{L_x}{2}\times L_y = \SI{2}{\meter}\times \SI{1}{\meter}$ is subject to its self-weight  and an external load, F$_\text{ext}$.  $L_x$ and $L_y$ indicate the length in $x-$ and $y-$directions respectively. Gravity load $\mathbf{g}$ is indicated by parallel arrows.}\label{fig:2DProblems}
\end{figure}

Herein, the efficacy and robustness of the proposed method are demonstrated by optimizing various structures with self-weight and/or constant external  loads. Both 2D and 3D designs are considered. The design domains with  boundary conditions are displayed in their respective section. $L_x$, $L_y$ and $L_z$ are used to indicate  length in $x-$, $y-$ and $z-$directions respectively. Thickness of the domain is represented by $t$ which is set to $\SI{0.01}{\meter}$ for the presented 2D design problems. We use $N_\text{ex}\times N_\text{ey}$ bilinear quadrilateral finite elements to describe the 2D design domains, whereas 3D domains are represented via $N_\text{ex}\times N_\text{ey} \times N_\text{ez}$ hexahedral FEs. $N_\text{ex}$, $N_\text{ey}$ and $N_\text{ez}$ denote the total number of FEs employed in $x-$, $y-$ and $z-$directions respectively to parameterize the design domain. One can also use honeycomb tessellation \citep{kumar2022honeytop90} to describe the design domain for 2D problems. A density-based TO approach is employed wherein each element is assigned one design variable which is considered constant within the element. TO process is initialized using the given volume fraction.  The Youngs' modulus and mass density of the material are set to $\SI{210}{GPa}$ and $\SI{7850}{\kilogram\per \meter\cubed}$ respectively  \citep{xu2013guide}. The total number of the MMA iterations is set to 250 unless  otherwise stated. As mentioned earlier, $\beta$ \eqref{eq:physicaldensity} is doubled after every 25 MMA iterations until it reaches its maximum value 256 and thereafter, it remains so for further iterations. We display the optimized material layouts of the solved numerical examples using their physical design vector $\bar{\mathbf{x}}$ \citep{sigmund2007morphology}. Note that design problems with self-weight have non-convex nature \citep{bruyneel2005note} and thus, we can neither ensure convergence to the global optimum nor avoid dependence of final solutions on the starting guesses.
\begin{figure}[h!]
	\begin{subfigure}[t]{0.22\textwidth}
		\begin{tikzpicture}
		\node[anchor=south west,inner sep=0] at (0,0){\includegraphics[scale=0.30]{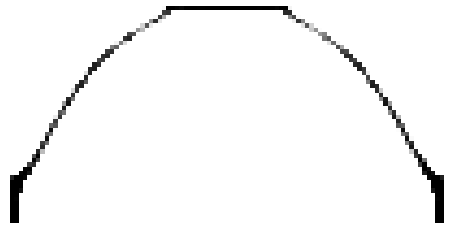}};
		\draw[red,line width = 1pt] (1.25,1.75)	circle (8pt);
		\draw[red,line width = 1pt] (2.5,1.75)	circle (8pt);
		\end{tikzpicture} 
		\caption{$f_0 = \SI{1.92e-2}{\newton\meter}$}
		\hspace*{30pt}	$V_f\mathbf{ =0.047}$
		\label{fig:CASE I} 
	\end{subfigure}
	\quad
	\begin{subfigure}[t]{0.22\textwidth}
		\centering
		\includegraphics[scale=0.30]{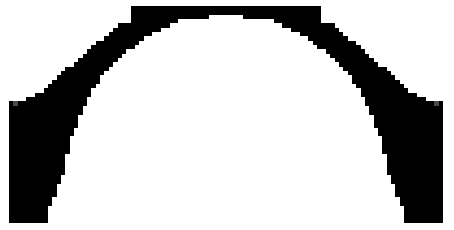}
		\caption{$f_0 =  \SI{1.2e-4}{\newton\meter}$}
		$V_f =0.25$
		\label{fig:CASE II}
	\end{subfigure}
	\quad 
	\begin{subfigure}[t]{0.22\textwidth}
		\centering
		\includegraphics[scale=0.30]{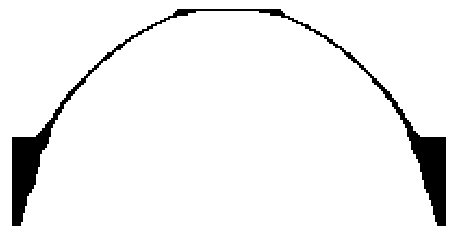}
		\caption{$f_0 =  \SI{0.2e-4}{\newton\meter}$}
		$V_f \mathbf{= 0.065}$
		\label{fig:CASE III}
	\end{subfigure}
	\quad 
	\begin{subfigure}[t]{0.22\textwidth}
		\centering
		\includegraphics[scale=0.30]{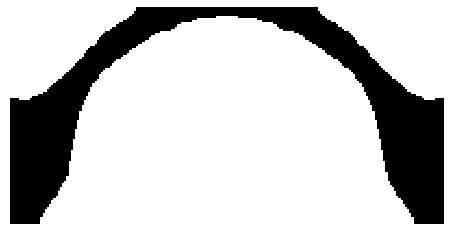}
		\caption{$f_0 = \SI{1.5e-4}{\newton\meter}$}
		$V_f = 0.25$
		\label{fig:CASE IV}
	\end{subfigure}
	\caption{The optimized results to the arch structure with various cases. (\subref{fig:CASE I}) CASE~I: $100 \times 50$ FEs without constraint g$_2$  (\subref{fig:CASE II}) CASE~II: $100 \times 50$ FEs  with constraint g$_2$,  (\subref{fig:CASE III})  CASE~III: $200 \times 100$ FEs  without constraint g$_2$, (\subref{fig:CASE IV}) CASE~IV: $200 \times 100$ FEs with constraint g$_2$. The optimized solution displayed in (\subref{fig:CASE I}) is disconnected as encircled in red. $V_f$ indicates the obtained volume fraction at the end of optimization. Optimization problem becomes unconstrained for CASE~I and CASE~II as their volume fractions are not active at the end of the optimization.}\label{fig:Example1_study_solutions}
\end{figure}

\begin{figure}[h!]
	\begin{subfigure}[t]{0.45\textwidth}
		\centering
		\begin{tikzpicture} 	
			\pgfplotsset{compat =1.9}
			\begin{axis}[
				width = 1\textwidth,
				xlabel=MMA iteration,
				ylabel= Volume fraction,
				legend style={at={(0.75,0.65)},anchor=east}]
				\pgfplotstableread{Qualifyvolf1.txt}\mydata;
				\addplot[smooth,red, mark size=2pt,style={very thick}, dotted]
				table {\mydata};
				\addlegendentry{CASE~I}
				\pgfplotstableread{Qualifyvolf2.txt}\mydata;
				\addplot[smooth,blue,mark size=1pt,style= thick]
				table {\mydata};
				\addlegendentry{CASE~II}
			\end{axis}
		\end{tikzpicture}
		\caption{}
		\label{fig:QualifyVolf1}
	\end{subfigure}
	\quad
	\begin{subfigure}[t]{0.45\textwidth}
		\centering
		\begin{tikzpicture}
			\pgfplotsset{compat = 1.9}
			\begin{axis}[
				width = 1\textwidth,
				xlabel=MMA iteration,
				ylabel= Volume fraction,
				legend style={at={(0.75,0.65)},anchor=east}]
				\pgfplotstableread{Qualifyvolf3.txt}\mydata;
				\addplot[smooth,red,mark size=1pt,style={very thick}, dotted]
				table {\mydata};
				\addlegendentry{CASE~III}
				\pgfplotstableread{Qualifyvolf4.txt}\mydata;
				\addplot[smooth,blue,mark size=1pt,style={thick}]
				table {\mydata};
				\addlegendentry{CASE~IV}
			\end{axis}
		\end{tikzpicture}
		\caption{}
		\label{fig:QualifyVolf2}
	\end{subfigure}
	\caption{Volume fraction convergence plots for CASE~I, CASE~II, CASE~III, CASE~IV.}
	\label{fig:InverterOConvergence}
\end{figure}
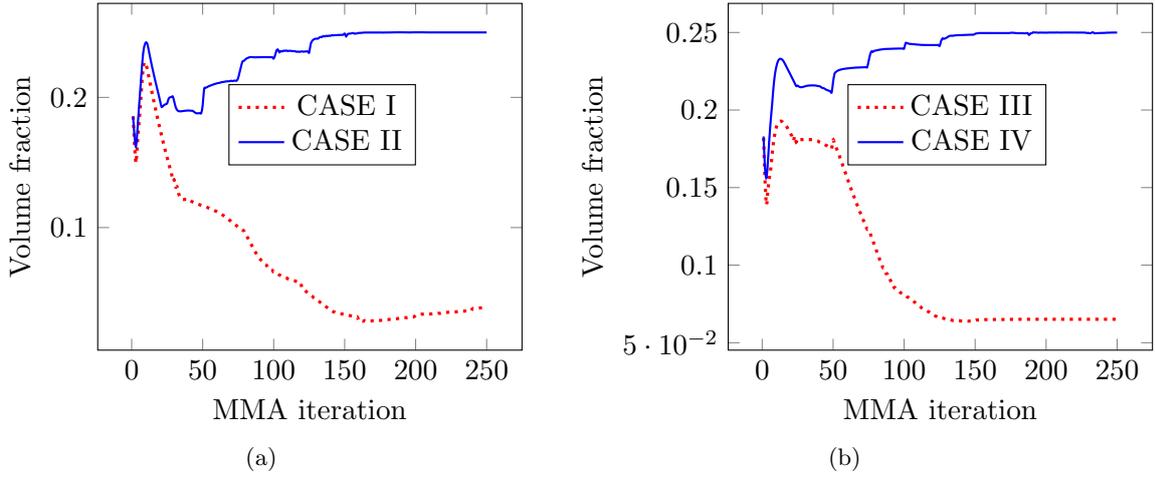

\begin{figure}[h!]
	\begin{subfigure}[t]{0.45\textwidth}
		\centering
		\begin{tikzpicture} 	
		\pgfplotsset{compat =1.9}
		\begin{axis}[
		width = 1\textwidth,
		xlabel=MMA iteration,
		ylabel= Constraints,
		legend style={at={(0.75,0.65)},anchor=east}]
		\pgfplotstableread{g1case2.txt}\mydata;
		\addplot[smooth,blue, mark size=2pt,style={very thick},dashed]
		table {\mydata};
		\addlegendentry{Constraint g$_1$}
		\pgfplotstableread{g2case2.txt}\mydata;
		\addplot[smooth,black, mark size=1pt,style= {thick}]
		table {\mydata};
		\addlegendentry{Constraint g$_2$}
		\end{axis}
		\end{tikzpicture}
		\caption{}
		\label{fig:case2constraints}
	\end{subfigure}
	\quad
	\begin{subfigure}[t]{0.45\textwidth}
		\centering
		\begin{tikzpicture}
		\pgfplotsset{compat = 1.9}
		\begin{axis}[
		width = 1\textwidth,
		xlabel=MMA iteration,
		ylabel= Constraints,
		legend style={at={(0.75,0.65)},anchor=east}]
		\pgfplotstableread{g1case4.txt}\mydata;
		\addplot[smooth,blue,mark size=1pt,style={very thick},,dashed]
		table {\mydata};
		\addlegendentry{Constraint g$_1$}
		\pgfplotstableread{g2case4.txt}\mydata;
		\addplot[smooth,black,mark size=1pt,style={thick}]
		table {\mydata};
		\addlegendentry{Constraint g$_2$}
		\end{axis}
		\end{tikzpicture}
		\caption{}
		\label{fig:case4constraints}
	\end{subfigure}
	\caption{Constraints convergence plots. (\subref{fig:case2constraints}) CASE~II and (\subref{fig:case4constraints}) CASE~IV.}
	\label{fig:constraintconvergence}
\end{figure}
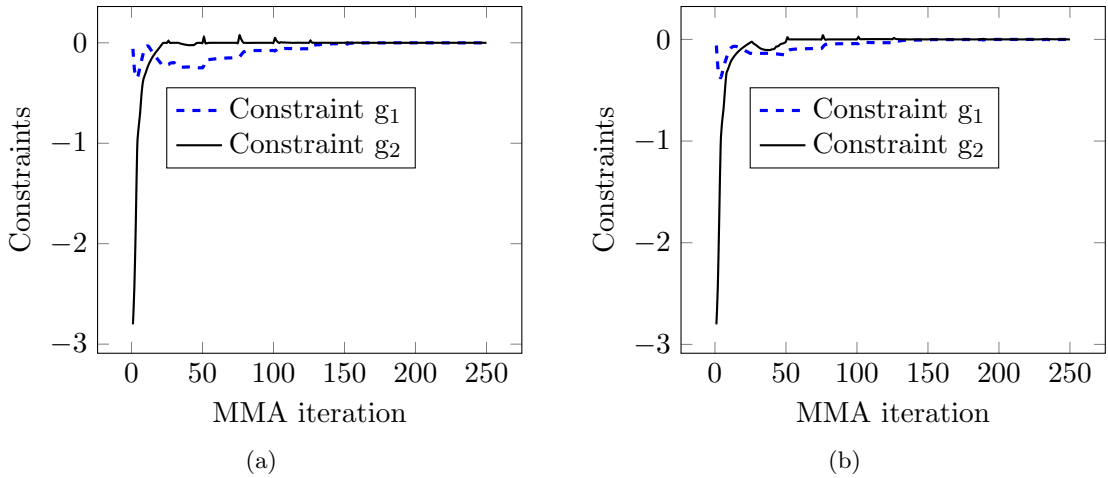
\subsection{2D design problems}
We first present various  2D  design-problems experiencing self-weight herein. 

\subsubsection{Self-weight loadbearing arch geometry}\label{sec:selfloadbearinggeometry}
The design domain for the self-weight loadbearing arch structure is depicted in Fig.~\ref{fig:problem 1}. Both ends of the bottom edge are fixed~(Fig.~\ref{fig:problem 1}), and no external force is applied (i.e $\kappa = 0$). $L_x = \SI{2}{\meter}$ and $L_y = \SI{1}{\meter}$ are taken. The external move limit of the optimizer is set to 0.1. Filter radius is set to $2.5\times \max\left(\frac{L_x}{N_\text{ex}},\frac{L_y}{N_\text{ey}}\right)$. The mass density parameters $\eta_\gamma = 0.01$ and $\beta_\gamma = 8$ are used \eqref{eq:Heavisidefunction}. The permitted volume fraction is set to 0.25.

\subsubsection*{Qualifying constraint $\mathrm{g}_2$}\label{sec:qualify g2}
	
Problems exclusively with self-weight (when self-weight effects dominate) under a volume constraint lose their constrained  nature \citep{bruyneel2005note}. To prevent this tendency,  additional constraint g$_2$ which implicitly ensures a lower bound on the resource volume is considered in conjunction with the presented mass density interpolation scheme~\eqref{eq:materialdensity} and the three-field representation technique~\citep{lazarov2016length}. The constraint is qualified and substantiated its importance by optimizing the arch design (Fig.~\ref{fig:problem 1}) with  self-weight \eqref{eq:Optimizationequation} using different cases and also, by optimizing the MBB beam design in the next section. 

To parameterize the design domain, (i) $N_\text{ex}\times N_\text{ey} = 100\times 50$ FEs and (ii) $N_\text{ex}\times N_\text{ey} = 200\times 100$ FEs are used. This is done to record the behavior of the approach with respect to different mesh sizes. In addition, though the problem is symmetric, the full model is considered to note any tendency to lose the symmetry in the optimized designs. Four cases namely, (i) CASE~I: $100 \times 50$ FEs without constraint g$_2$, (ii) CASE~II: $100 \times 50$ FEs with constraint g$_2$, (iii)  CASE~III: $200 \times 100$ FEs  without constraint g$_2$, and (iv) CASE~IV: $200 \times 100$ FEs with constraint g$_2$ are considered.

\begin{figure}[h!]
	\centering
	\begin{subfigure}[t]{0.45\textwidth}
		\centering
		\includegraphics[scale=0.65]{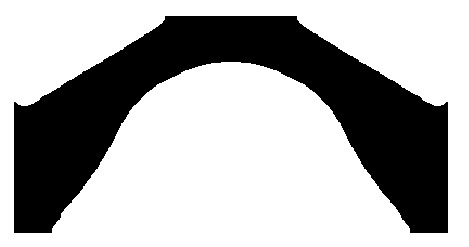}
		\caption{Optimized arch structure}
		\label{fig:archsolution}
	\end{subfigure}
	\quad
	\begin{subfigure}[t]{00.45\textwidth}
		\centering
		\begin{tikzpicture}
			\pgfplotsset{compat = 1.9}
			\begin{axis}[
				width = 1\textwidth,
				xlabel=MMA iteration,
				ylabel= Volume fraction,
				legend style={at={(0.75,0.65)},anchor=east}]
				\pgfplotstableread{archvolf.txt}\mydata;
				\addplot[smooth,blue,mark = *,mark size=1pt,style={thick}]
				table {\mydata};
			\end{axis}
		\end{tikzpicture}
		\caption{}
		\label{fig:archvolumconv}
	\end{subfigure}
	\caption{(\subref{fig:archsolution}) Optimized arch design, $V_f = 0.40,\,f_0 = \SI{4.7e-4}{\newton \meter}$, (\subref{fig:archvolumconv}) Volume fraction convergence curve }
\end{figure}

\begin{figure}
	\centering
	\includegraphics[scale=2.0]{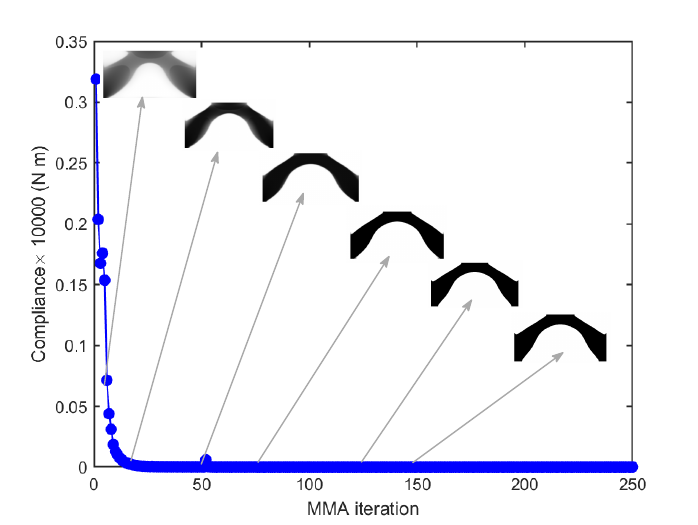}
	\caption{Objective convergence history of the arch structure with intermediate results.}
	\label{fig:archobjconv}
\end{figure}

\begin{figure}[h!]
	\centering
	\begin{subfigure}[t]{0.30\textwidth}
		\centering
		\includegraphics[scale=0.25]{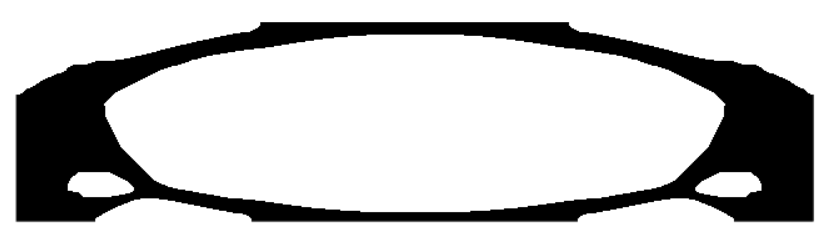}
		\caption{$V^*_f = 0.3$}
		\label{fig:Ex2volf0.3}
	\end{subfigure}
	\quad
	\begin{subfigure}[t]{00.30\textwidth}
		\centering
		\includegraphics[scale=0.13]{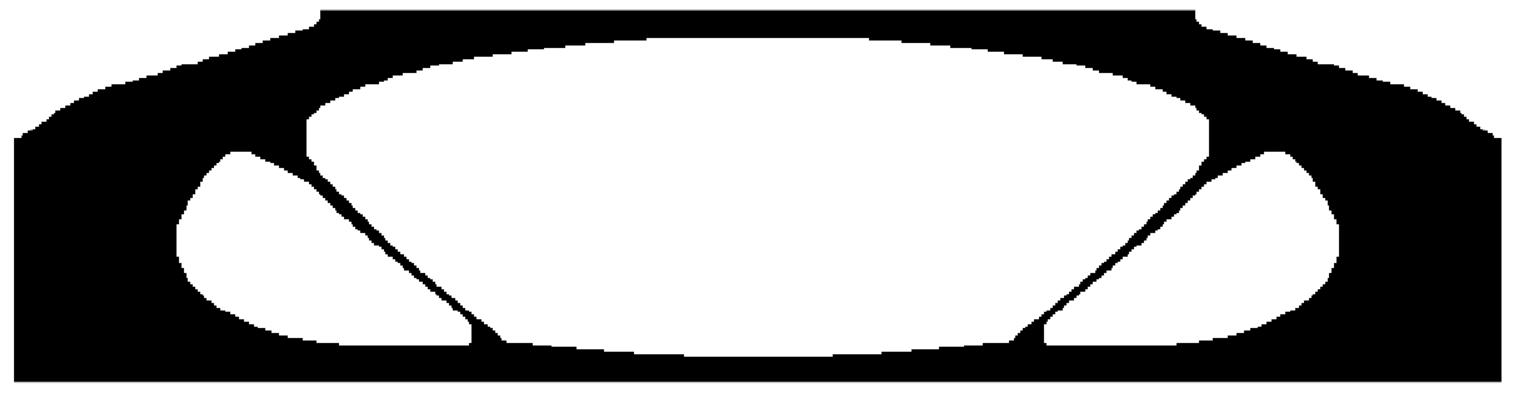}
		\caption{$V^*_f = 0.4$}
		\label{fig:Ex2volf0.4}
	\end{subfigure}
	\quad
	\begin{subfigure}[t]{00.30\textwidth}
		\centering
		\includegraphics[scale=0.13]{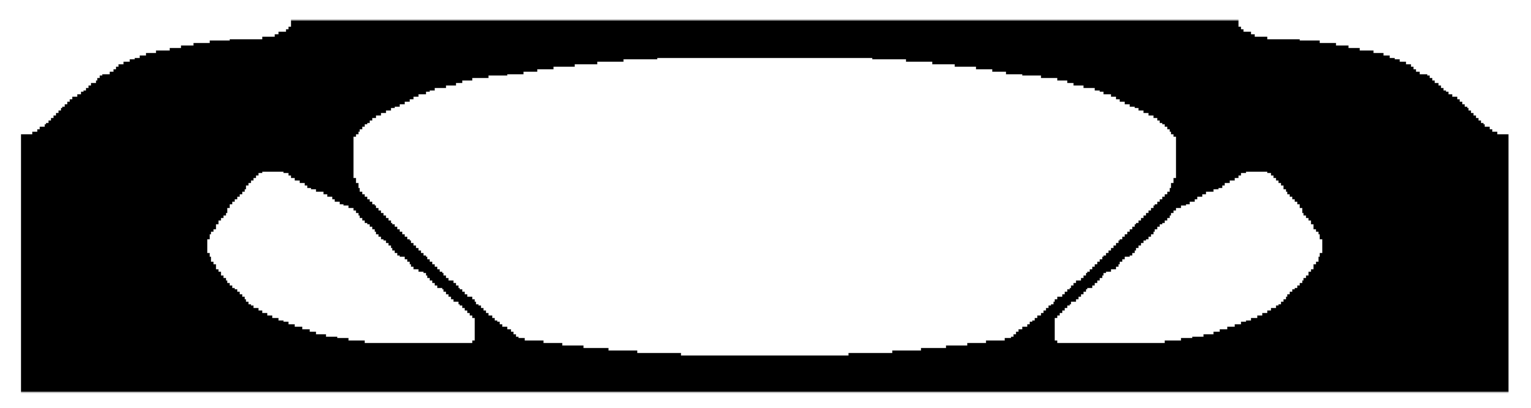}
		\caption{$V^*_f = 0.5$}
		\label{fig:Ex2volf0.5}
	\end{subfigure}
	\caption{ Optimized designs for the MBB beam design}
\end{figure}

The optimized results for all the cases are displayed  after 250 MMA iterations in Fig.~\ref{fig:Example1_study_solutions}. When constraint g$_2$ is not taken into account, for example in CASE~I and CASE~III, the optimization problem becomes unconstrained as evident from their volume constraint convergence plots (see Fig.~\ref{fig:QualifyVolf1} and Fig.~\ref{fig:QualifyVolf2}). The optimized design of CASE~I contains gray elements (marked in red circles in Fig.~\ref{fig:CASE I}). On the other hand, with constraint g$_2$ (CASE~II and CASE~IV) the optimization problem retains its constrained nature and at the end of TO,  volume constraint~ g$_1$ stays active (Fig.~\ref{fig:QualifyVolf1} and Fig.~\ref{fig:QualifyVolf2}). Steps in volume fraction convergence curves are due to $\beta$  updation scheme \eqref{eq:physicaldensity} which is performed to achieve 0-1 optimized solutions. Constraints' convergence plots for CASE~II and CASE~IV are displayed in Fig.~\ref{fig:case2constraints} and Fig.~\ref{fig:case4constraints} respectively. Constraints are active at the end of optimization. The optimized designs of CASE~II~(Fig.~\ref{fig:CASE II}) and CASE~IV~(Fig.~\ref{fig:CASE IV}) are close to 0-1 solutions. Topologies of these solutions resemble those reported in \citep{bruyneel2005note,huang2011evolutionary,novotny2021topological}. Furthermore, these solutions (Fig.~\ref{fig:CASE II} and Fig.~\ref{fig:CASE IV}) are symmetric with respect to the central $y-$axis of the domain and thus, symmetry nature of the problem is retained.  We henceforth report convergence history for constraint g$_1$ only in view of Fig.~\ref{fig:constraintconvergence}.

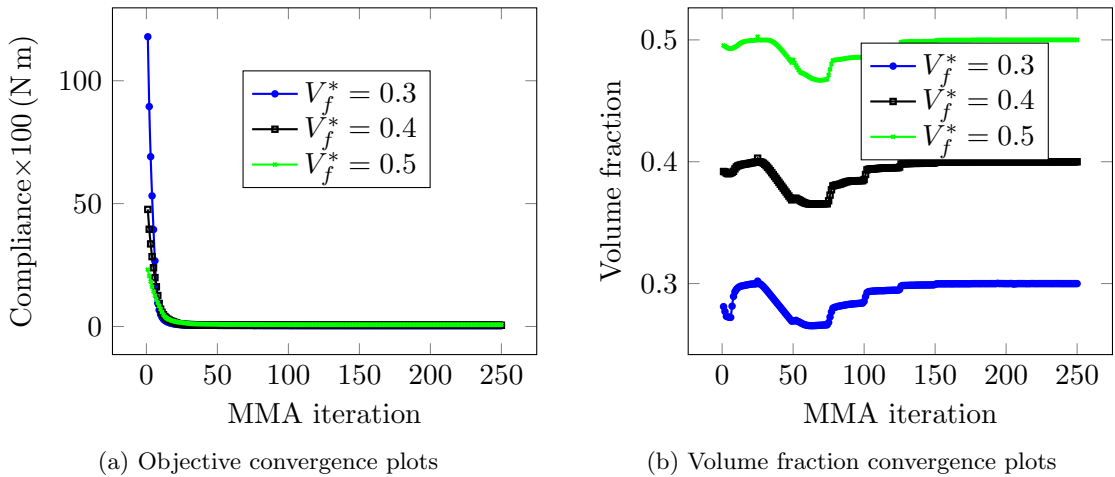
\begin{figure}[h!]
	\begin{subfigure}[t]{00.45\textwidth}
		\centering
		\begin{tikzpicture}
			\pgfplotsset{compat = 1.9}
			\begin{axis}[
				width = 1\textwidth,
				xlabel=MMA iteration,
				ylabel= Compliance$\times 100\,(\si{\newton \meter})$ ,
				legend style={at={(0.75,0.65)},anchor=east}]
				\pgfplotstableread{ex3selfvf03obj.txt}\mydata;
				\addplot[smooth,blue,mark = *,mark size=1pt,style={thick}]
				table {\mydata};
				\addlegendentry{$V^*_f = 0.3$}
				\pgfplotstableread{ex3selfvf04obj.txt}\mydata;
				\addplot[smooth,black,mark = square,mark size=1pt,style={thick}]
				table {\mydata};
				\addlegendentry{$V^*_f = 0.4$}
				\pgfplotstableread{ex3selfvf05obj.txt}\mydata;
				\addplot[smooth,green,mark = x,mark size=1pt,style={thick}]
				table {\mydata};
				\addlegendentry{$V^*_f = 0.5$}
			\end{axis}
		\end{tikzpicture}
		\caption{Objective convergence plots}
		\label{fig:Ex2objconv}
	\end{subfigure}
	\quad
	\begin{subfigure}[t]{00.45\textwidth}
		\centering
		\begin{tikzpicture}
			\pgfplotsset{compat = 1.9}
			\begin{axis}[
				width = 1\textwidth,
				xlabel=MMA iteration,
				ylabel= Volume fraction,
				legend style={at={(0.85,0.73)},anchor=east}]
				\pgfplotstableread{ex3selfvf03volf.txt}\mydata;
				\addplot[smooth,blue,mark = *,mark size=1pt,style={thick}]
				table {\mydata};
				\addlegendentry{$V^*_f = 0.3$}
				\pgfplotstableread{ex3selfvf04volf.txt}\mydata;
				\addplot[smooth,black,mark = square,mark size=1pt,style={thick}]
				table {\mydata};
				\addlegendentry{$V^*_f = 0.4$}
				\pgfplotstableread{ex3selfvf05volf.txt}\mydata;
				\addplot[smooth,green,mark = x,mark size=1pt,style={thick}]
				table {\mydata};
				\addlegendentry{$V^*_f = 0.5$}
			\end{axis}
		\end{tikzpicture}
		\caption{Volume fraction convergence plots}
		\label{fig:Ex2volfconv}
	\end{subfigure}
	\caption{Converge plots for the compliance and volume fractions.}
\end{figure}

\begin{figure}[h!]
	\centering
	\begin{subfigure}[t]{0.30\textwidth}
		\centering
		\includegraphics[scale=0.45]{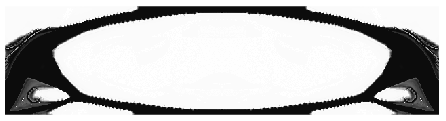}
		\caption{}
		\label{fig:originalfield}
	\end{subfigure}
	\quad
	\begin{subfigure}[t]{00.30\textwidth}
		\centering
		\includegraphics[scale=0.45]{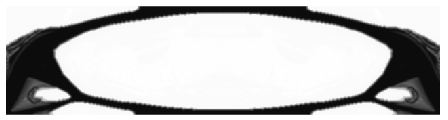}
		\caption{}
		\label{fig:filterfield}
	\end{subfigure}
	\quad
	\begin{subfigure}[t]{00.30\textwidth}
		\centering
		\includegraphics[scale=0.45]{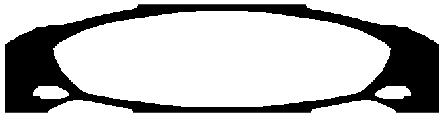}
		\caption{}
		\label{fig:projected field}
	\end{subfigure}
	\caption{Final material layouts with original, filtered and physical fields of the optimized result displayed in Fig.~\ref{fig:Ex2volf0.3} are shown in (\subref{fig:originalfield}), (\subref{fig:filterfield}) and (\subref{fig:projected field}) respectively.}\label{fig:orifilprojectedfields}
\end{figure}

\subsubsection{Arch structure}\label{sec:archstructure}
The design domain (Fig.~\ref{fig:problem 1}) is described using $N_\text{ex}\times N_\text{ey} = 400\times 200$ FEs. The volume fraction is set to~0.40. The external move limit of the MMA is set to 0.05 herein and used henceforth. $\eta_\gamma = 0.1$ and $\beta_\gamma = 8$ are considered. Filter radius is set to $3.5\max\left({\frac{L_x}{N_\text{ex}}},\,\frac{L_y}{N_\text{ey}}\right)$. Constraint g$_2$ is considered within the optimization formulation. Other design parameters are same as those employed in Sec.~\ref{sec:selfloadbearinggeometry}.

The optimized result is depicted in Fig.~\ref{fig:archsolution}, and the corresponding convergence curves for the  volume fraction the objective are displayed in  Fig.~\ref{fig:archvolumconv} and Fig.~\ref{fig:archobjconv} respectively. The design evolution at different intermediate stages are also shown in Fig.~\ref{fig:archobjconv}. The objective convergence is smooth and relatively rapid. This implies that no unbounded displacements exist and thus, parasitic effects of the low-stiffness elements are circumvented.  As $\beta$ increases, the boundaries of the solution become crisp, and the solution moves towards 0-1 design (Fig.~\ref{fig:archobjconv}). The volume constraint is active and satisfied at the end of optimization and thus, constrained nature of the problem is maintained. As $\beta$ \eqref{eq:physicaldensity} is updated at every 25 MMA iterations, corresponding steps in volume fraction can be  noted~(Fig.~\ref{fig:archvolumconv}).

\begin{figure}[h!]
	\centering
	\begin{subfigure}[t]{0.225\textwidth}
		\centering
		\includegraphics[scale=0.3]{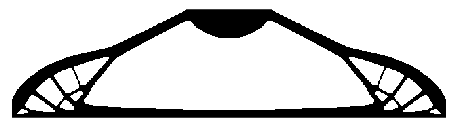}
		\caption{$\kappa = 3$ and self-weight}
		\label{fig:Ex2_ex_3_self_1}
	\end{subfigure}
	\quad
	\begin{subfigure}[t]{00.225\textwidth}
		\centering
		\includegraphics[scale=0.3]{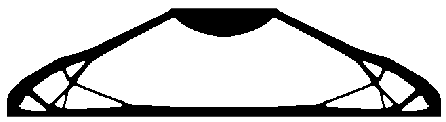}
		\caption{$\kappa =2$ and self-weight}
		\label{fig:Ex2_ex_2_self_1}
	\end{subfigure}
	\quad
	\begin{subfigure}[t]{0.225\textwidth}
		\centering
		\includegraphics[scale=0.3]{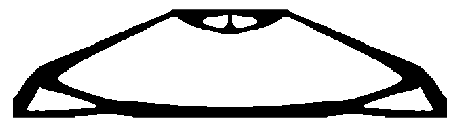}
		\caption{$\kappa = 1$ and self-weight}
		\label{fig:Ex2_ex_1_self_1}
	\end{subfigure}
	\quad
	\begin{subfigure}[t]{00.225\textwidth}
		\centering
		\includegraphics[scale=0.3]{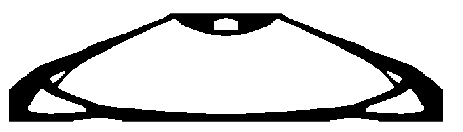}
		\caption{$\kappa =0.75$ and self-weight}
		\label{fig:Ex2_ex_075_self_1}
	\end{subfigure}
	\quad
	\begin{subfigure}[t]{0.225\textwidth}
		\centering
		\includegraphics[scale=0.3]{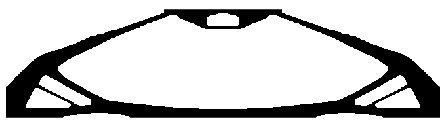}
		\caption{$\kappa = 0.5$ and self-weight}
		\label{fig:Ex2_ex_05_self_1}
	\end{subfigure}
	\quad
	\begin{subfigure}[t]{0.225\textwidth}
		\centering
		\includegraphics[scale=0.3]{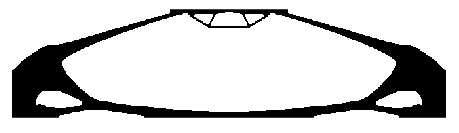}
		\caption{$\kappa = 0.25$ and self-weight}
		\label{fig:Ex2_ex_025_self_1}
	\end{subfigure}
	\quad
	\begin{subfigure}[t]{0.225\textwidth}
		\centering
		\includegraphics[scale=0.3]{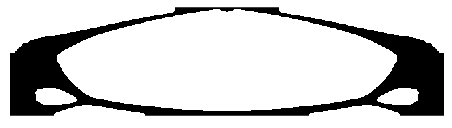}
		\caption{$\kappa = 0.1$ and self-weight}
		\label{fig:Ex2_ex_01_self_1}
	\end{subfigure}
	\quad
	\begin{subfigure}[t]{0.225\textwidth}
		\centering
		\includegraphics[scale=0.3]{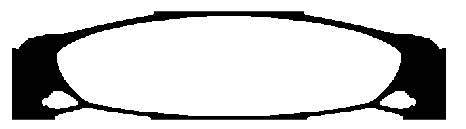}
		\caption{$\kappa = 0$ and self-weight}
		\label{fig:Ex2_ex_0_self_1}
	\end{subfigure}
	\caption{Optimized results to the MBB beam \textit{with constraint} g$_2$ for different $\kappa$. (\subref{fig:Ex2_ex_3_self_1}) $f_0 = \SI{15.90e-2}{\newton \meter}$, (\subref{fig:Ex2_ex_2_self_1}) $f_0 = \SI{8.90e-2}{\newton \meter}$, (\subref{fig:Ex2_ex_1_self_1}) $f_0 = \SI{3.83e-2}{\newton \meter}$, (\subref{fig:Ex2_ex_075_self_1}) $f_0 = \SI{2.64e-2}{\newton \meter}$, (\subref{fig:Ex2_ex_05_self_1}) $f_0 = \SI{1.73e-2}{\newton \meter}$,   (\subref{fig:Ex2_ex_025_self_1}) $f_0 = \SI{0.83e-2}{\newton \meter}$,  (\subref{fig:Ex2_ex_01_self_1}) $f_0 = \SI{0.45e-2}{\newton \meter}$,  (\subref{fig:Ex2_ex_0_self_1}) $f_0 = \SI{0.24e-2}{\newton \meter}$. $V_f \mathbf{= 0.25}$ is noted for all the results displayed here.} \label{fig:Ex2kappa}
\end{figure}

 \begin{figure}[h!]
	\centering
	\begin{subfigure}[t]{0.225\textwidth}
		\centering
		\includegraphics[scale=0.3]{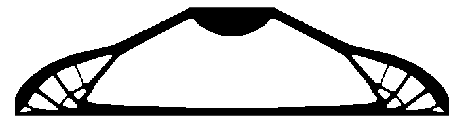}
		\caption{$\kappa = 3$ and self-weight}
		\footnotesize	$V_f =0.25$
		\label{fig:MBBL3}
	\end{subfigure}
	\quad
	\begin{subfigure}[t]{.225\textwidth}
		\centering
		\includegraphics[scale=0.3]{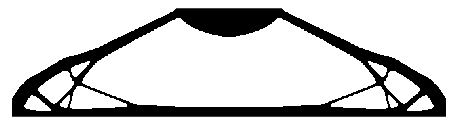}
		\caption{$\kappa =2$ and self-weight}
		\footnotesize	$V_f\mathbf{ =0.247}$
		\label{fig:MBBL2}
	\end{subfigure}
	\quad
	\begin{subfigure}[t]{0.225\textwidth}
		\centering
		\includegraphics[scale=0.3]{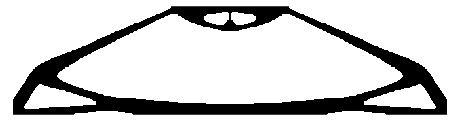}
		\caption{$\kappa = 1$ and self-weight}
		\footnotesize	$V_f \mathbf{=0.238}$
		\label{fig:MBBL1}
	\end{subfigure}
	\quad
	\begin{subfigure}[t]{0.23\textwidth}
		\centering
		\includegraphics[scale=0.3]{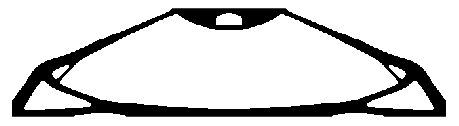}
		\caption{$\kappa = 0.75$ and self-weight}
		\footnotesize	$V_f \mathbf{=0.22}$
		\label{fig:MBBL4}
	\end{subfigure}
	\quad
	\begin{subfigure}[t]{0.225\textwidth}
		\centering
		\includegraphics[scale=0.3]{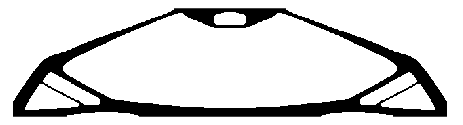}
		\caption{$\kappa = 0.5$ and self-weight}
		\footnotesize	$V_f\mathbf{ =0.20}$
		\label{fig:MBBL5}
	\end{subfigure}
	\quad
	\begin{subfigure}[t]{0.225\textwidth}
		\centering
		\includegraphics[scale=0.3]{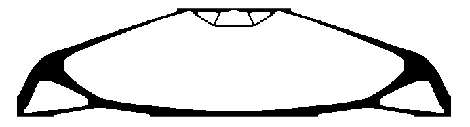}
		\caption{$\kappa = 0.25$ and self-weight}
		\footnotesize	$V_f \mathbf{=0.17}$
		\label{fig:MBBL6}
	\end{subfigure}
	\quad
	\begin{subfigure}[t]{0.225\textwidth}
		\centering
		\includegraphics[scale=0.3]{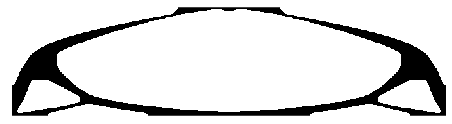}
		\caption{$\kappa = 0.1$ and self-weight}
		\footnotesize	$V_f \mathbf{=0.16}$
		\label{fig:MBBL7}
	\end{subfigure}
	\quad
	\begin{subfigure}[t]{0.23\textwidth}
		\centering
		\includegraphics[scale=0.3]{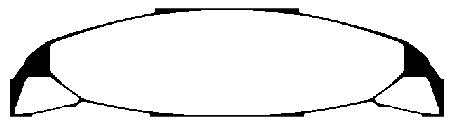}
		\caption{$\kappa = 0.25$ and self-weight}
		\footnotesize	$V_f \mathbf{=0.098}$
		\label{fig:MBBL8}
	\end{subfigure}
	\caption{Optimized results to the MBB beam \textit{without constraint} g$_2$ for different $\kappa$. (\subref{fig:MBBL3}) $f_0 = \SI{15.90e-2}{\newton \meter}$, (\subref{fig:MBBL2}) $f_0 = \SI{8.93e-2}{\newton \meter}$, (\subref{fig:MBBL1}) $f_0 = \SI{3.92e-2}{\newton \meter}$,   (\subref{fig:MBBL4}) $f_0 = \SI{2.78e-2}{\newton \meter}$,  (\subref{fig:MBBL5}) $f_0 = \SI{1.87e-2}{\newton \meter}$,  (\subref{fig:MBBL6}) $f_0 = \SI{0.86e-2}{\newton \meter}$, (\subref{fig:MBBL7}) $f_0 = \SI{0.43e-2}{\newton \meter}$, (\subref{fig:MBBL8}) $f_0 = \SI{0.935e-3}{\newton \meter}$. $V_f$ denotes the final volume fraction.} \label{fig:Ex2kappawithoutg2}
\end{figure}

\subsubsection{MBB beam design}\label{sec:MBBbeamdesign}

A Messerschmitt-Bolkow-Blohm (MBB) beam design subjected to self-weight  as well as an external load is studied in this example. 

In view of symmetric nature of the problem, the right symmetric part  (Fig.~\ref{fig:problem 2}) with dimension  $\frac{L_x}{2} = \SI{2}{\meter}$ and $L_y = \SI{1}{\meter}$ is considered. Magnitude of the external load, applied as shown in Fig.~\ref{fig:problem 2} in the negative $y-$direction, is set to $\text{F}_\text{g}^{\text{max}} = V\gamma_sV_f^*\text{g}$. Different $\kappa$ is taken herein. The boundary conditions of the problem are as depicted in Fig.~\ref{fig:problem 2}.  The domain is parameterized using $N_\text{ex}\times N_\text{ey} = 320\times 160$ FEs. The filter radius is set to $3\times \max\left(\frac{L_x}{N_\text{ex}},\frac{L_y}{N_\text{ey}}\right)$. Other parameters are same as those used in Sec.~\ref{sec:selfloadbearinggeometry}. In this study, three cases are considered.

First, the beam is optimized with $\kappa = 0$, i.e, only self-weight  and  with different volume fractions, e.g., $V^*_f = 0.3,\,0.4,\,0.5$. The problem is solved with constraint g$_2$. The optimized results are displayed in Figs.~\ref{fig:Ex2volf0.3}-\ref{fig:Ex2volf0.5}, which are very close to 0-1 with similar topologies. Each optimized design has a horizontal bar-typed slender member which is expected with such boundary conditions.  As  volume fraction increases more material get accumulated near the support in the optimized structures (Figs.~\ref{fig:Ex2volf0.3}-\ref{fig:Ex2volf0.5}). This occurs to reduce the bending moment of self-weight of the material placed by the optimizer. Convergence curves of the objectives and the volume fractions are depicted in Fig.~\ref{fig:Ex2objconv} and Fig.~\ref{fig:Ex2volfconv} respectively.  The objective convergence is smooth and rapid (Fig.~\ref{fig:Ex2objconv}) that is also noted previously. Parasitic effects of the low-stiffness elements/regions are not observed. Constrained nature of the problem is preserved as the respective volume constraint is active at the end of optimization (Fig.~\ref{fig:Ex2volfconv}). Moreover, the trends of volume fraction convergence plots are similar as the employed  $\beta$ continuation is same for them. The steps in volume convergence curves (Fig.~\ref{fig:Ex2volfconv}) are associated with $\beta$ updation. Figure~\ref{fig:orifilprojectedfields} displays the material layout plots of original (Fig.~\ref{fig:originalfield}), filtered (Fig.~\ref{fig:filterfield}) and physical (Fig.~\ref{fig:projected field}) fields of the optimized result shown in Fig.~\ref{fig:Ex2volf0.3}. Fig.~\ref{fig:originalfield}, Fig.~\ref{fig:filterfield} and Fig.~\ref{fig:projected field} have  similar typologies. Gray elements can be noted in Fig.~\ref{fig:originalfield} and Fig.~\ref{fig:filterfield}, however the result in Fig.~\ref{fig:projected field} (Fig.~\ref{fig:Ex2volf0.3}) is close to 0-1.

\begin{figure}[h!]
	\centering
	\begin{subfigure}[t]{0.25\textwidth}
		\centering
		\includegraphics[scale=1.0]{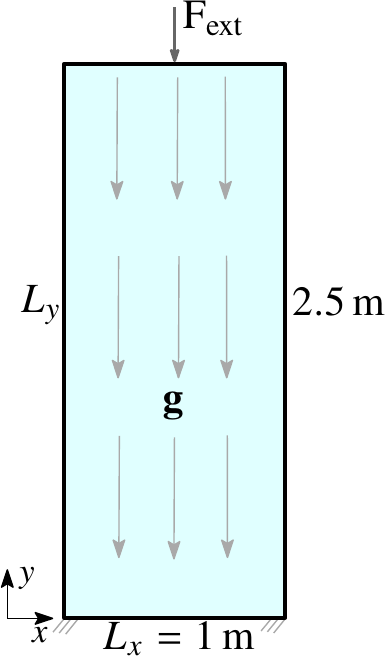}
		\caption{Tower design domain}
		\label{fig:problem 3}
	\end{subfigure}
	\begin{subfigure}[t]{0.25\textwidth}
		\centering
		\includegraphics[scale=0.64]{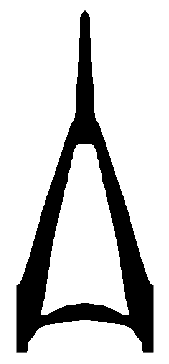}
		\caption{$f_0 = \SI{2.24e-3}{\newton\meter}$}
		\label{fig:Ex3solumaterial}
	\end{subfigure}
	\begin{subfigure}[t]{0.45\textwidth}
		\centering
		\begin{tikzpicture}
			\pgfplotsset{compat = 1.3}
			\begin{axis}[
				width = 1\textwidth,
				xlabel=MMA iteration,
				axis y line* = left,
				ylabel= Compliance $\times 100\,(\si{\newton \meter})$]
				\pgfplotstableread{Ex3objconv.txt}\mydata;
				\addplot[smooth,blue,mark = *,mark size=1pt,style={thick}]
				table {\mydata};\label{plot1}
			\end{axis}
			\begin{axis}[
				width = 1\textwidth,
				xlabel=MMA iteration,
				axis y line* = right,
				ylabel= Volume fraction,
				legend style={at={(0.95,0.5)},anchor=east}]
				\addlegendimage{/pgfplots/refstyle=plot1}\addlegendentry{Objective}
				\pgfplotstableread{Ex3volfconv.txt}\mydata;
				\addplot[smooth,black,mark = square,mark size=1pt,style={thick}]
				table {\mydata};
				\addlegendentry{volume fraction}
			\end{axis}
		\end{tikzpicture}
		\caption{Convergence plots}
		\label{fig:Ex3objvolfconv}
	\end{subfigure}
	\caption{(\subref{fig:problem 3}) Design domain of dimension $L_x\times L_y = \SI{1}{\meter}\times \SI{2.5}{\meter}$ is under its self-weight  and  external load F$_\text{ext}$, (\subref{fig:Ex3solumaterial}) Optimized design, the final volume $V_f = 0.25$ is obtained, (\subref{fig:Ex3objvolfconv}) Convergence curves for the objective and volume fraction.}	\label{fig:Ex3solu}
\end{figure}
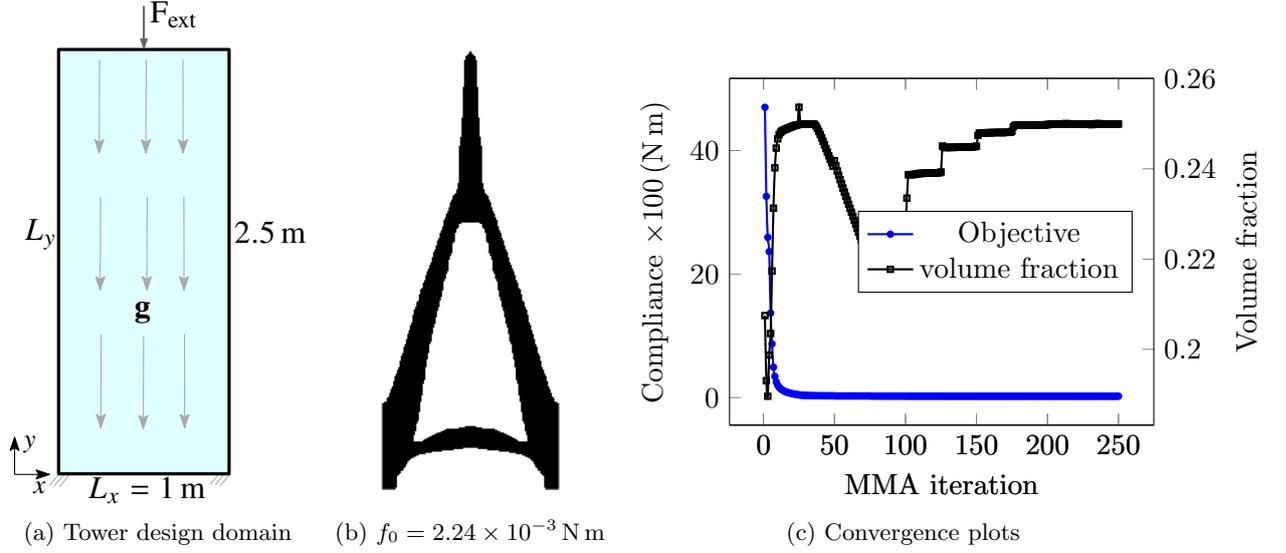

Second, we solve the MBB beam design with different magnitudes of F$_\text{ext}$, i.e., with different $\kappa$ and constraint g$_2$. The desired resource volume is set to $25\%$. Other parameters are same as above. The optimized results  are shown in Fig.~\ref{fig:Ex2kappa}. One can note that as $\kappa$ decreases, the material gets transferred from the center region to the lateral sides of the domain so that the optimized designs experience less bending moment due to the self-weight. The central structures which support the external load disappear eventually as magnitude of F$_\text{ext}$ decreases. The geometry of the optimized layout gradually changes to two connected arch structures facing towards each other. The final compliance decreases as $\kappa$ decreases that is as expected. The convergence of objective is found to be smooth and rapid. The volume fractions are satisfied and remain active for all $\kappa$.

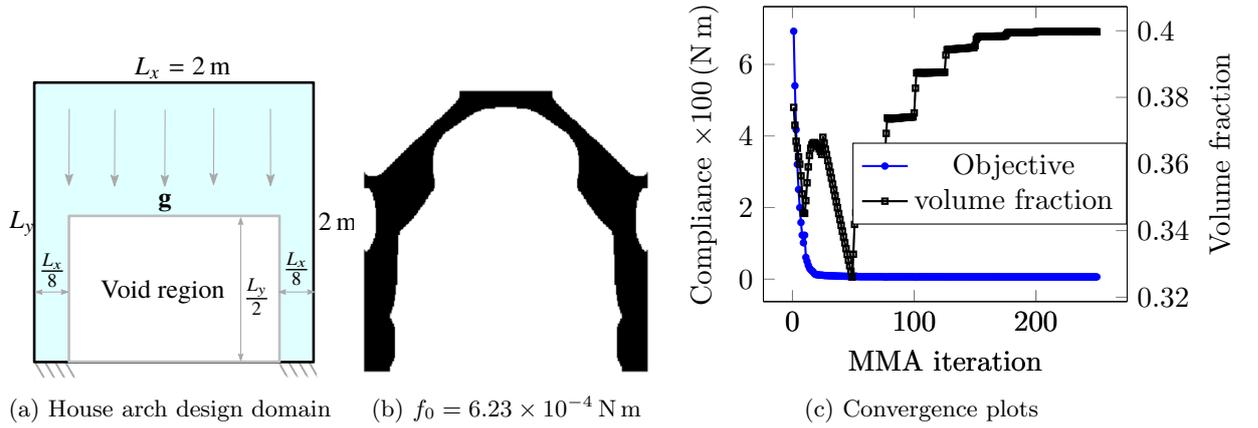
\begin{figure}[h!]
	\centering
	\begin{subfigure}[t]{0.27\textwidth}
		\centering
		\includegraphics[scale=0.82]{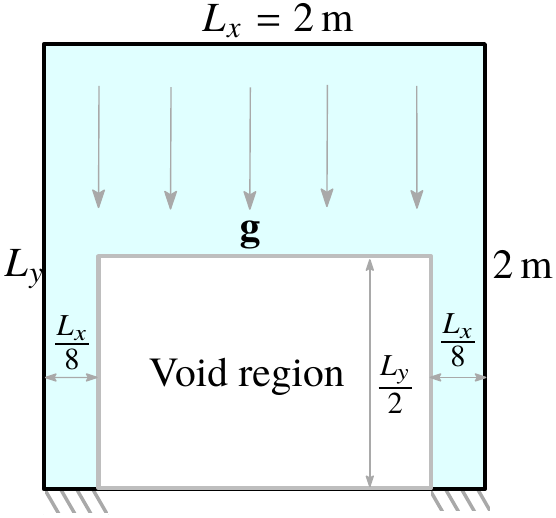}
		\caption{House arch design domain}
		\label{fig:problem 4}
	\end{subfigure}
	\begin{subfigure}[t]{0.27\textwidth}
		\centering
		\includegraphics[scale=0.41]{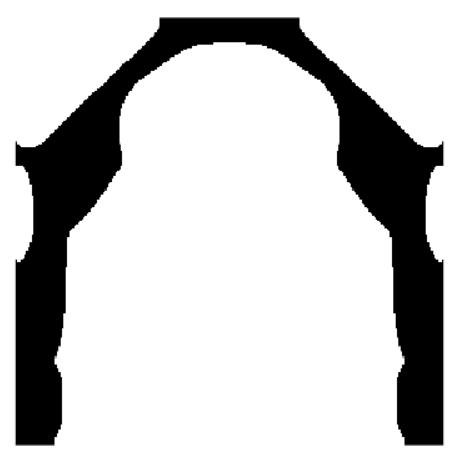}
		\caption{$f_0 = \SI{6.23e-4}{\newton\meter}$}
		\label{fig:Ex4solumaterial}
	\end{subfigure}
	\begin{subfigure}[t]{0.40\textwidth}
		\centering
		\begin{tikzpicture}
		\pgfplotsset{compat = 1.3}
		\begin{axis}[
		width = 1\textwidth,
		xlabel=MMA iteration,
		axis y line* = left,
		ylabel= Compliance $\times 100\,(\si{\newton \meter})$]
		\pgfplotstableread{Ex4objconv.txt}\mydata;
		\addplot[smooth,blue,mark = *,mark size=1pt,style={thick}]
		table {\mydata};\label{plot1.1}
		\end{axis}
		\begin{axis}[
		width = 1\textwidth,
		xlabel=MMA iteration,
		axis y line* = right,
		ylabel= Volume fraction,
		legend style={at={(1.0,0.40)},anchor=east}]
		\addlegendimage{/pgfplots/refstyle=plot1.1}\addlegendentry{Objective}
		\pgfplotstableread{Ex4volfconv.txt}\mydata;
		\addplot[smooth,black,mark = square,mark size=1pt,style={thick}]
		table {\mydata};
		\addlegendentry{volume fraction}
		\end{axis}
		\end{tikzpicture}
		\caption{Convergence plots}
		\label{fig:Ex4objvolfconv}
	\end{subfigure}
	\caption{(\subref{fig:problem 3}) Design domain of dimension $L_x\times L_y = \SI{2}{\meter}\times \SI{2}{\meter}$ is subject to its self-weight. A non-design void region of dimension $\frac{7 L_x}{4}\times \frac{L_y}{2}$ is present as depicted (\subref{fig:Ex3solumaterial}) Optimized design, (\subref{fig:Ex3objvolfconv}) Convergence curves for the objective and volume fraction.}	\label{fig:Ex4solu}
\end{figure}

Third, the MBB beam design problem is solved without constraint g$_2$ for different $\kappa$ (Fig.~\ref{fig:Ex2kappawithoutg2}). The corresponding optimized results are displayed in Fig.~\ref{fig:Ex2kappawithoutg2}. We note that for $\kappa\le 2$ (Figs.~\ref{fig:MBBL2}-\subref{fig:MBBL8})  the volume constraint of the problem  is although satisfied, is not active at the end of optimization, i.e, the constrained nature of the problem is not retained. On the other hand, the final material volume with $\kappa = 3$ is 0.25, i.e., the volume constraint remains active at the end of optimization. Therefore,  mere presence of an external load with self-weight for a design problem cannot help retain the constrained nature of the problem unless the magnitude of the applied  external load is relatively large~(Fig.~\ref{fig:MBBL3}) or a lower bound on the permitted material volume is employed (implicitly/explicitly) ~(Fig.~\ref{fig:Ex2kappa}). In other words, as long as the effects of self-weight are prominent in the optimization process, we need constraint like g$_2$ within the optimization formulation to retain the constrained nature of the problem. We henceforth solve all problems with constraint g$_2$.

\subsubsection{Tower design}
For this example, a design domain for tower structure is considered with self-weight  and an external load.

The design domain is displayed in Fig.~\ref{fig:problem 3}. Both ends of the bottom edge are fixed.  $L_x\times L_y = \SI{1}{\meter}\times \SI{2.5}{\meter}$ is taken. Filter radius is equal to $5.6\times \max\left(\frac{L_x}{N_\text{ex}},\frac{L_y}{N_\text{ey}}\right)$. In light of the vertical symmetry, only a symmetric half design domain is consider for simulation and optimization.  $110 \times 550$ FEs are employed to parameterize the symmetric half domain. Volume fraction is set to 0.25. Magnitude of the external load is taken equal to the self-weight. The load is applied at the center of the top edge of domain in the negative $y-$direction. Other design parameters are same as above.

The optimized tower structure is shown in Fig.~\ref{fig:Ex3solumaterial}. The objective and the volume fraction convergence plots are displayed in Fig.~\ref{fig:Ex3objvolfconv}. A rapid and smooth convergence for the objective can be noted. The parasitic effects of the low-stiffness regions do not exist. The volume fraction is satisfied and remains active at the end of optimization. The optimized structure is close to 0-1. A vertical slender structure appears in the optimized design (Fig.~\ref{fig:Ex3solumaterial}) to contain the applied external load. By and large, the optimized design resembles a typical tower design.

\begin{figure}[h!]
	\centering
	\includegraphics[scale=1.85]{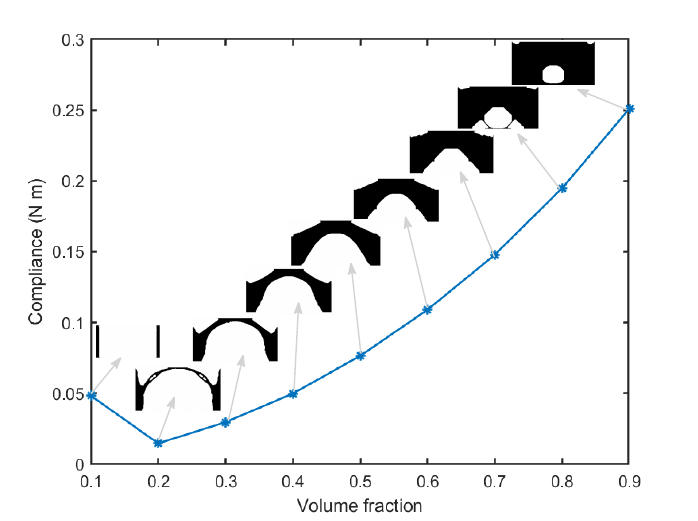}
	\caption{A Pareto curve and optimized arch designs for different volume fraction} \label{fig:volumefractions}
\end{figure}
\begin{figure}[h!]
	\centering
	\begin{subfigure}[t]{0.22\textwidth}
		\centering
		\includegraphics[scale=0.30]{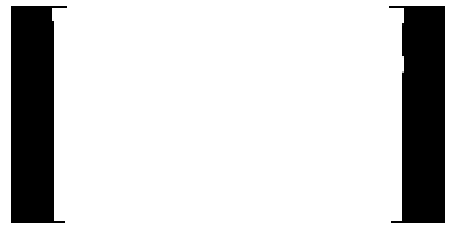}
		\caption{$f_0 = \SI{0.8823e-2}{\newton \meter}$}
		\label{fig:MD1}
	\end{subfigure}
	\quad
	\begin{subfigure}[t]{0.22\textwidth}
		\centering
		\includegraphics[scale=0.30]{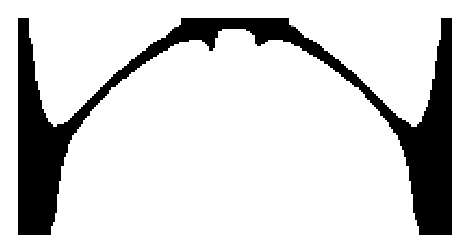}
		\caption{$f_0 = \SI{0.0155e-2}{\newton \meter}$}
		\label{fig:MD2}
	\end{subfigure}
	\quad
	\begin{subfigure}[t]{0.22\textwidth}
		\centering
		\includegraphics[scale=0.30]{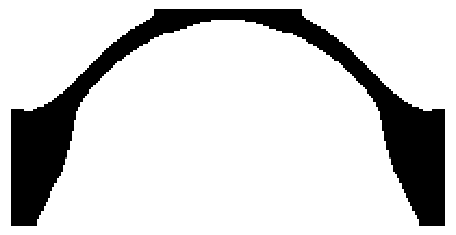}
		\caption{$f_0 = \SI{0.0143e-2}{\newton \meter}$}
		\label{fig:MD3}
	\end{subfigure}
	\quad
	\begin{subfigure}[t]{0.22\textwidth}
		\centering
		\includegraphics[scale=0.30]{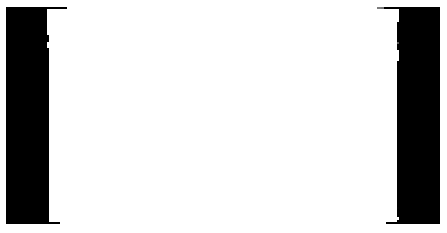}
		\caption{$f_0 = \SI{0.8820e-2}{\newton \meter}$}
		\label{fig:MD4}
	\end{subfigure}
	\caption{Optimized results. (\subref{fig:MD1})$\eta_\gamma = 0.1,\,\beta_\gamma = 10$, (\subref{fig:MD2}) $\eta_\gamma = 0.05,\,\beta_\gamma = 10$, (\subref{fig:MD3})$\eta_\gamma = 0.01,\,\beta_\gamma = 10$,   and (\subref{fig:MD4}) $\eta_\gamma = 0.1,\,\beta_\gamma = 20$} \label{fig:MD}
\end{figure}

\subsubsection{House arch design}
In this example, the presented approach is demonstrated by designing a structure resembling a typical house arch structure. 

The design domain and boundary conditions are shown in Fig.~\ref{fig:problem 4} wherein $L_x \times L_y = \SI{2}{\meter}\times \SI{2}{\meter}$. A non-design void region of size $\frac{7 L_x}{4}\times \frac{L_y}{2}$~(Fig.~\ref{fig:problem 4}) is present in the design domain to facilitate entry and exit. The structure is considered under self-weight  only, i.e., $\kappa=0$. Though a vertical symmetry exists, we take full domain to analyze and optimize so that any deviation from symmetry can be noted in presence of a non-design domain. The domain is parameterized by $N_\text{ex}\times N_\text{ey} = 240\times 240$ FEs. The permitted volume fraction is 0.40. Filter radius is set to $3.6\times \max\left(\frac{L_x}{N_\text{ex}},\,\frac{L_y}{N_\text{ey}}\right)$. Other parameters are same as those used above.

The optimized result is displayed in Fig.~\ref{fig:Ex4solumaterial}. The optimized design is constituted via two pillars and an arch structure on the top. The corresponding convergence plots are depicted in Fig.~\ref{fig:Ex4objvolfconv}. As noted earlier here too the objective convergence is rapid and smooth. One can note that the volume constraint is active at the end of optimization (Fig.~\ref{fig:Ex4objvolfconv}). The steps in volume fraction convergence curve are due to $\beta$ updation. This indicates that method works well when non-design domains are present. 

\begin{figure}[h!]
		\centering
	\begin{subfigure}[t]{0.22\textwidth}
		\centering
		\includegraphics[scale=0.30]{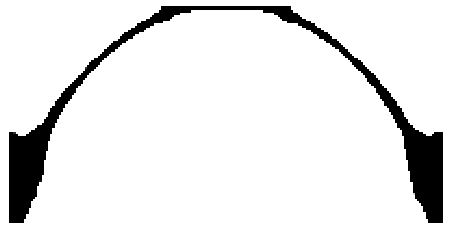}
		\caption{$V^*_f = 0.1$}
		\quad \quad $\eta_\gamma = 0.001,\, \beta_\gamma = 20$
		\label{fig:vf01MD}
	\end{subfigure}
	\quad
	\begin{subfigure}[t]{0.22\textwidth}
		\centering
		\includegraphics[scale=0.30]{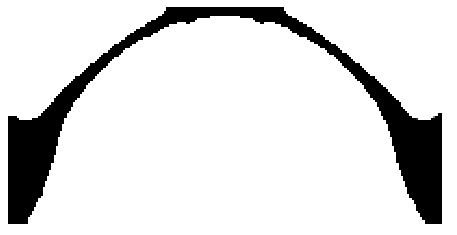}
		\caption{$V_f^* = 0.15$}
		\quad \quad $\eta_\gamma = 0.003,\, \beta_\gamma = 12$
		\label{fig:vf015MD}
	\end{subfigure}
	\quad
	\begin{subfigure}[t]{0.22\textwidth}
		\centering
		\includegraphics[scale=0.30]{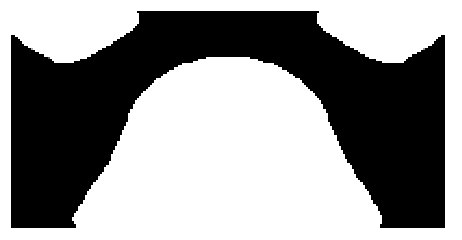}
		\caption{$V_f^* = 0.50$}
		\quad \quad $\eta_\gamma = 0.15,\, \beta_\gamma = 6$
		\label{fig:vf05MD}
	\end{subfigure}
	\quad
	\begin{subfigure}[t]{0.22\textwidth}
		\centering
		\includegraphics[scale=0.30]{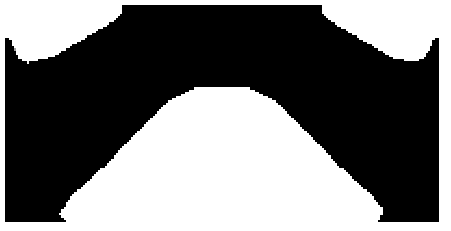}
		\caption{$V_f^* = 0.6$}
		\quad \quad $\eta_\gamma = 0.225,\, \beta_\gamma = 8$
		\label{fig:vf06MD}
	\end{subfigure}
	\caption{Optimized results. (\subref{fig:vf01MD})$f_0 = \SI{4.6e-5}{\newton \meter}$, (\subref{fig:vf015MD}) $f_0 = \SI{8.7e-5}{\newton \meter}$, (\subref{fig:vf05MD}) $f_0 = \SI{7.69e-4}{\newton \meter}$,   and (\subref{fig:vf06MD}) $f_0 = \SI{1.1e-3}{\newton \meter}$} \label{fig:MDsuggest}
\end{figure}

\subsection{Parameter study}\label{sec:paramterstudy}
In this section, we present the effects of different parameters on the optimized designs with self-weight.
\subsubsection{Volume fraction}
Herein, different volume fractions ranging from 0.1 to 0.9 are used to solve the arch structure (Fig.~\ref{fig:problem 1}). $N_\text{ex}\times N_\text{ey} = 200 \times 100$ FEs is used to represent the design domain. $\beta_\gamma = 6$ and $\eta_\gamma = 0.01$ are considered.  Filter radius is set to $2.5\times \max\left(\frac{L_x}{N_\text{ex}},\,\frac{L_y}{N_\text{ey}}\right)$. Other design parameters are same as those used in Sec.~\ref{sec:selfloadbearinggeometry}.

Fig.~\ref{fig:volumefractions} displays a Pareto curve between compliance and  volume fractions. As volume fraction increases, self-weight of the optimized design increases and thus, corresponding compliance increases. The optimized results are shown in Fig.~\ref{fig:volumefractions} for every volume fraction. It is noted that the volume constraint for each case remains active at the end of optimization, and the corresponding objective convergence is rapid and smooth.  The optimized design with volume fraction 0.1 is disconnected, which is constituted via two pillars (see Sec.~\ref{sec:Massdensityparameters}). Compliance with $V_f^* = 0.1$ is obtained higher than that with  $V_f^* = 0.2$.

\subsubsection{Mass density parameters}\label{sec:Massdensityparameters}
This study demonstrates the effects of different mass density parameters $\{\eta_\gamma,\,\beta_\gamma\}$ on the optimized designs with self-weight. 

The design domain for this study is shown in Fig.~\ref{fig:problem 1}. Volume fraction $V^*_f = 0.20$ is set, and filter radius is taken equal to  $2.5\times \max\left(\frac{L_x}{N_\text{ex}},\,\frac{L_y}{N_\text{ey}}\right)$.  $N_\text{ex}\times N_\text{ey} = 200 \times 100$ FEs is employed to describe the design domain. Note that, $\beta_\gamma$ controls the slope of material density interpolation (Fig.~\ref{fig:Materialdensity}). For higher $\beta_\gamma$, elements with $\bar{{x}} \ge \eta_\gamma$ act as solid FEs.

The optimized designs with various $\left\{\eta_\gamma,\,\beta_\gamma\right\}$ are depicted in Fig.~\ref{fig:MD}. Topologies of the optimized designs are similar in Fig.~\ref{fig:MD2} and Fig.~\ref{fig:MD3} but having different final objective values. The obtained final objective value with $\{\eta_\gamma,\,\beta_\gamma \} = \{0.01,\,20\}$ is lower than those of all other cases considered, which suggests and confirms that indeed an arch-shaped structure is the actual optimized design for the problem shown in Fig.~\ref{fig:problem 1}.  Optimized designs shown in Fig.~\ref{fig:MD1}, Fig.~\ref{fig:MD4} are disconnected, and their final compliance are higher than the other two in Fig.~\ref{fig:MD}. The optimized design displays in Fig.~\ref{fig:MD2} is asymmetric (central region). A possible reason may be the numerical noise during optimization. Note however that with a suitable choice of $\{\eta_\gamma,\,\beta_\gamma\}$, one can also retain symmetric nature of the problems in their optimized designs (Fig.~\ref{fig:MD3}) as progress of the optimization process is relatively stable i.e less noisy due to lower non-monotonous behavior of the objective. For the tuning the latter, $\eta_\gamma$ should be sufficiently far and towards left on design variable axis from the given volume fraction (Fig.~\ref{fig:Materialdensity}).  When decreasing $\eta_\gamma$ and increasing $\beta_\gamma$, the region of transitions from void to solid reduces (Fig.~\ref{fig:Materialdensity}) and in those cases, the optimized designs are found to be connected and sensible. Relatively moderate $\beta_\gamma$ and $\eta_\gamma$ offers smoother optimization problem, however the final obtained designs may be disconnected. Thus, in general, $\beta_\gamma$ and $\eta_\gamma$ can be chosen such that a suitable trade-off between the differentiability and transition i.e a suitable span of transition from void to solid phases for the given problem can be obtained (Fig.~\ref{fig:Materialdensity}) and also, non-monotonous behavior of the objective is reduced (Fig.~\ref{fig:self-weightderivative}). By and large, based on our experience, $\eta_\gamma$ less than and equal to ${(V^*_f)}^p$ and $\beta_\gamma$ between $5-20$ ensure the desired trade-off. To demonstrate that indeed the recommended values work, we solve the problem for permitted volume fractions 0.1, 0.15, 0.5 and 0.6 with $\{0.001,\,20\},\,\{0.003,\,12\},\,\{0.15,\,6\},\,\{0.225,\,8\}$ as their $\{\eta_\gamma,\,\beta_\gamma \}$. Fig.~\ref{fig:MDsuggest} shows the optimized design. These designs are connected and sensible. We also notice that choosing  $\eta_\gamma= 0$ and $\chi=0$ work fine, for all the cases reported with the respective $\beta_\gamma$, which simplifies while retaining the novelties of the proposed  mass density interpolation scheme~\eqref{eq:materialdensity} to $\gamma_e = \frac{\gamma_s\tanh(\beta_\gamma\bar{x}_e)}{\tanh (\beta_\gamma)}$ or $\gamma_e = \gamma_s\tanh(\beta_\gamma\bar{x}_e)$. In addition, the volume constraints for all cases are found to be satisfied and active at the end of optimization. 

\begin{figure}[h!]
	\begin{subfigure}[t]{0.48\textwidth}
		\centering
		\includegraphics[scale=1.25]{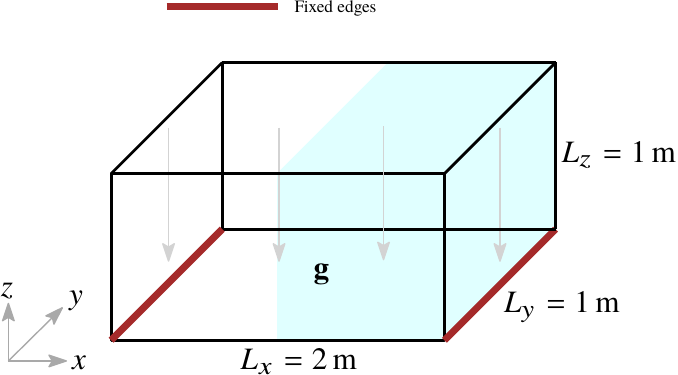}
		\caption{}
		\label{fig:3DP1}
	\end{subfigure}
	\begin{subfigure}[t]{0.48\textwidth}
		\centering
		\includegraphics[scale=1.25]{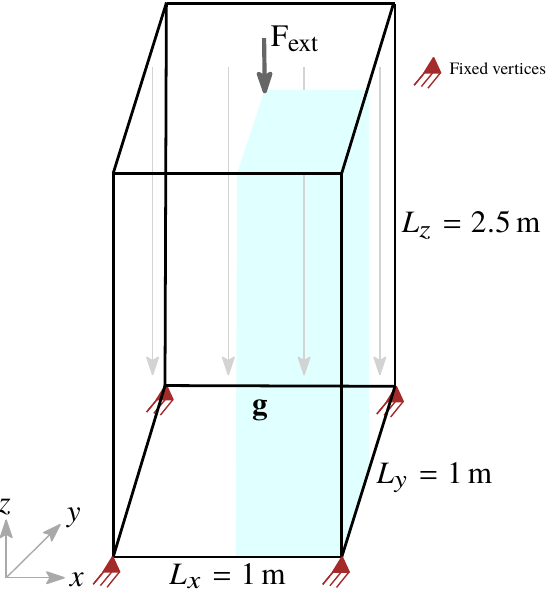}
		\caption{}
		\label{fig:3DP2}
	\end{subfigure}
	\caption{3D design domains. (\subref{fig:3DP1}) Arch structure domain and (\subref{fig:3DP2}) Tower design domain. Gravity is indicated via gray parallel arrows pointing in the negative $z-$direction. }
	\label{fig:3D design domain}
\end{figure}

\begin{figure}[h!]
		\centering
	\begin{subfigure}[t]{0.18\textwidth}
		\centering
		\includegraphics[scale=0.18]{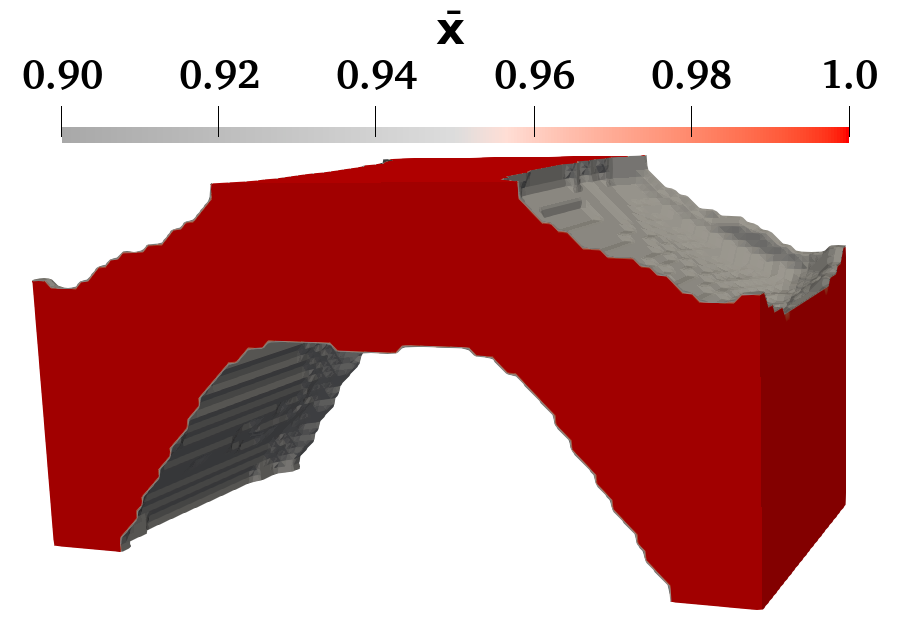}
		\caption{Isoview direction}
		\label{fig:3DarchV1}
	\end{subfigure}
	\,
	\begin{subfigure}[t]{0.18\textwidth}
		\centering
		\includegraphics[scale=0.18]{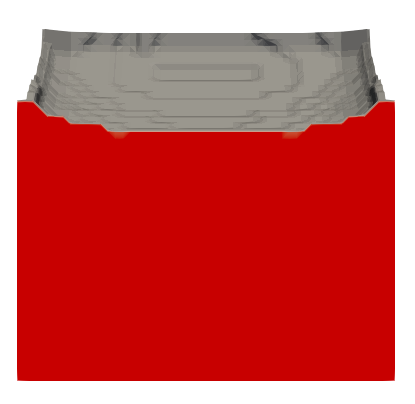}
		\caption{$+x-$direction}
		\label{fig:3Darchx+}
	\end{subfigure}
	\,
	\begin{subfigure}[t]{0.18\textwidth}
		\centering
		\includegraphics[scale=0.18]{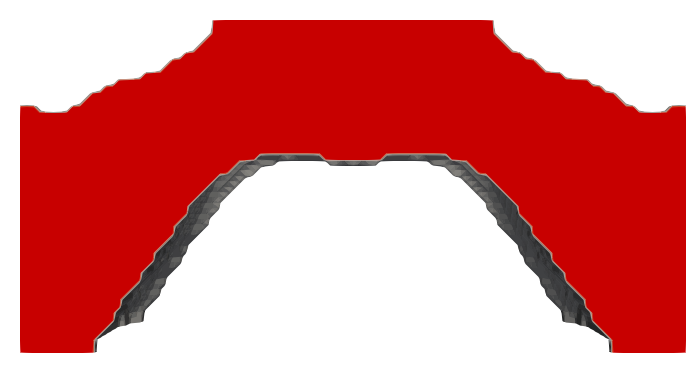}
		\caption{$+y-$direction}
		\label{fig:3Darchy+}
	\end{subfigure}
	\,
	\begin{subfigure}[t]{0.18\textwidth}
		\centering
		\includegraphics[scale=0.18]{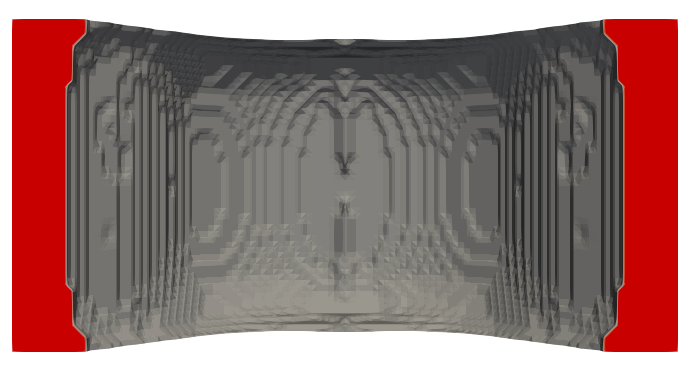}
		\caption{$+z-$direction}
		\label{fig:3Darchz+}
	\end{subfigure}
	\,
	\begin{subfigure}[t]{0.18\textwidth}
		\centering
		\includegraphics[scale=0.18]{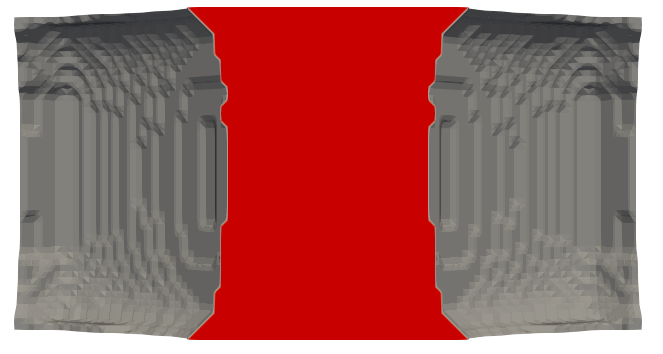}
		\caption{$-z-$direction}
		\label{fig:3Darchz-}
	\end{subfigure}
	\caption{3D Optimized results for the arch geometry are shown in different view directions.} \label{fig:3Darchresult}
\end{figure}

\begin{figure}[h!]
	\centering
	\begin{subfigure}[t]{0.30\textwidth}
		\centering
		\includegraphics[scale=0.40]{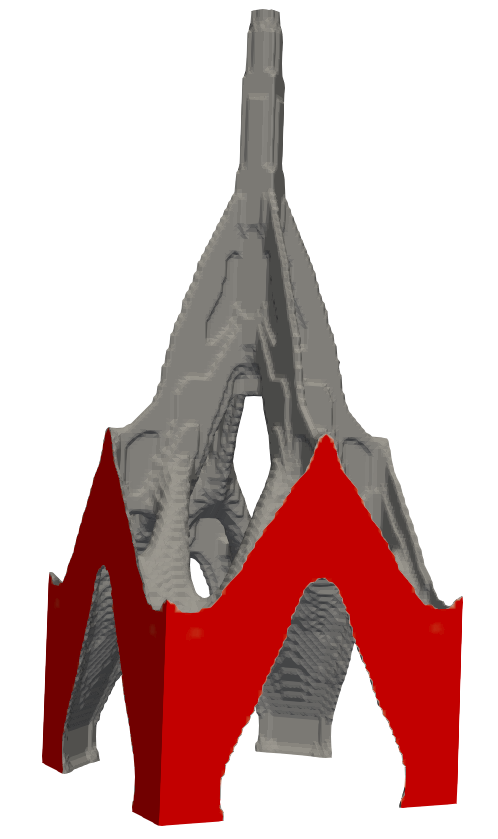}
		\caption{Isoview direction}
		\label{fig:3DTV1}
	\end{subfigure}
	\quad
	\begin{subfigure}[t]{0.30\textwidth}
		\centering
		\includegraphics[scale=0.40]{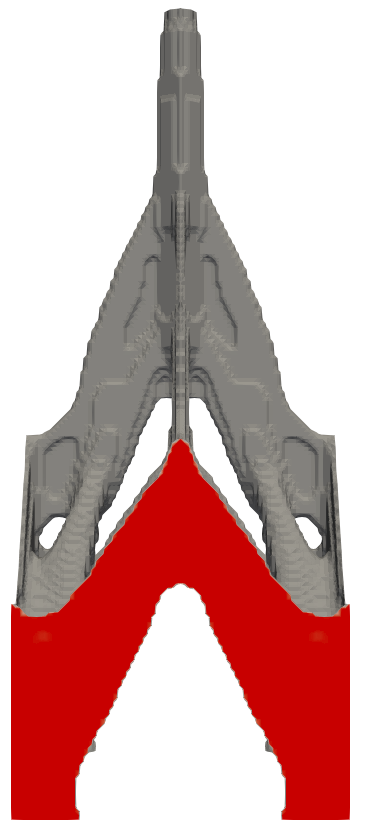}
		\caption{$+x-$direction}
		\label{fig:3DTVx+}
	\end{subfigure}
	\quad
	\begin{subfigure}[t]{0.30\textwidth}
		\centering
		\includegraphics[scale=0.40]{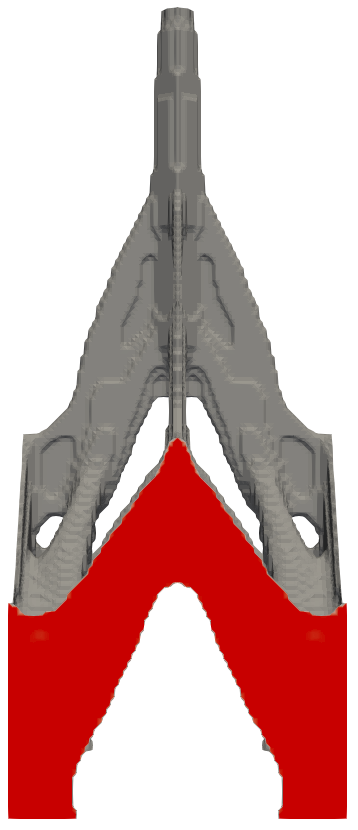}
		\caption{$+y-$direction}
		\label{fig:3DTVy+}
	\end{subfigure}
	\quad
	\begin{subfigure}[t]{0.45\textwidth}
		\centering
		\includegraphics[scale=0.50]{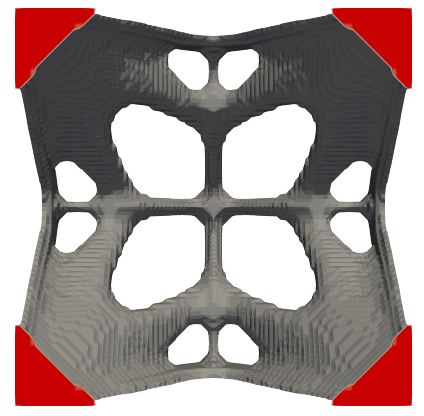}
		\caption{$+z-$direction}
		\label{fig:3DTVz+}
	\end{subfigure}
	\quad
	\begin{subfigure}[t]{0.45\textwidth}
		\centering
		\includegraphics[scale=0.65]{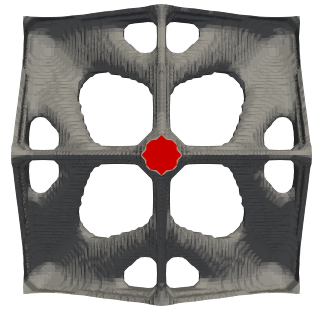}
		\caption{$-z-$direction}
		\label{fig:3DTVz-}
	\end{subfigure}
	\caption{3D Optimized results for a tower structure are displayed in different view directions.} \label{fig:3Dtowerresult}
\end{figure}

\begin{figure}[h!]
	\centering
	\begin{subfigure}[t]{0.45\textwidth}
		\centering
		\includegraphics[scale=0.60]{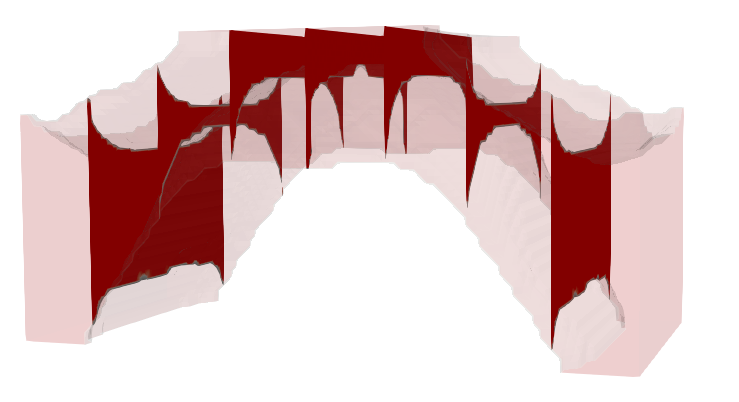}
		\caption{}
		\label{fig:3Darchslice}
	\end{subfigure}
	\quad
	\begin{subfigure}[t]{0.45\textwidth}
		\centering
		\includegraphics[scale=0.350]{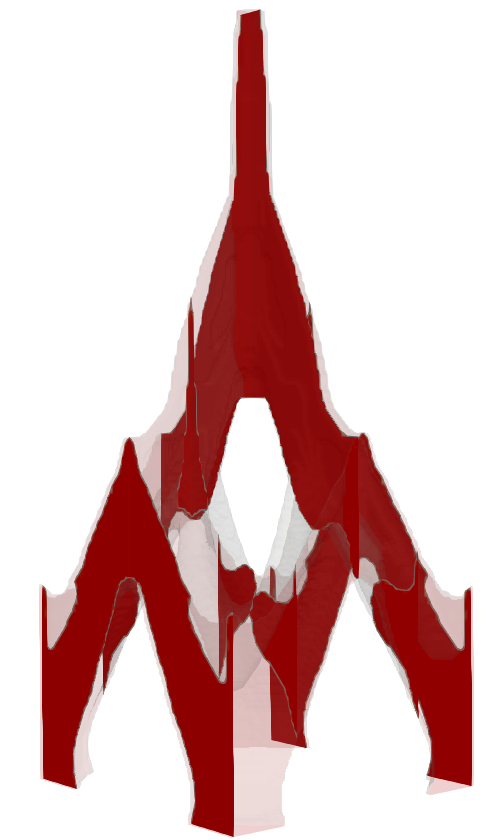}
		\caption{}
		\label{fig:3Dtowerslice}
	\end{subfigure}
	\caption{  Material distributions for different cross-sections in arbitrary directions (\subref{fig:3Darchslice}) Arch structure (\subref{fig:3Dtowerslice}) Tower structure.} \label{fig:3Darchtowerslice}
\end{figure}

\subsection{Three-dimensional examples}\label{sec:theedimension}

This section demonstrates that the proposed approach can without much difficulty be extended for 3D problems including self-weight.  TO is performed using an in-house MATLAB code wherein the conjugate gradient method in association with incomplete Cholesky preconditioning is employed to solve the linear systems from the equilibrium and adjoint equations.

Two 3D problems are solved: (i) an arch structure problem with only self-weight  and (ii)  a tower structure with self-weight  and a central constant load. Their design domains are displayed in Fig.~\ref{fig:3D design domain}. For the arch structure, we exploit one symmetry axis for the analysis and optimization (Fig.~\ref{fig:3DP1}), whereas for the tower problem both symmetry axes are exploited (Fig.~\ref{fig:3DP2}). The symmetric parts used for optimization are shadowed using light cyan color (Fig.~\ref{fig:3D design domain}).

$L_x\times L_y \times L_z = \SI{2}{\meter}\times \SI{1}{\meter}\times \SI{1}{\meter}$ and  $L_x\times L_y \times L_z = \SI{1}{\meter}\times \SI{1}{\meter}\times \SI{2.5}{\meter}$ are considered for the arch design (Fig.~\ref{fig:3DP1}) and the tower design (Fig.~\ref{fig:3DP2}) respectively. The half symmetric part of arch domain (colored shape in Fig.~\ref{fig:3DP1}) is parameterized via $N_\text{ex}\times N_\text{ey}\times N_\text{ez} = 50 \times 50 \times 50$ hexahedral FEs, whereas the quarter part of the  tower domain (colored shape in Fig.~\ref{fig:3DP2}) is described via $N_\text{ex}\times N_\text{ey}\times N_\text{ez} = 40\times 40 \times 200$ FEs. The permitted volume fractions,  filter radii and material density parameters  $\left\{\eta_\gamma,\,\beta_\gamma\right\}$ for  arch and tower design problems are set to 0.35 and 0.1, $4.8\times \max\left(\frac{L_x}{N_\text{ex}},\,\frac{L_y}{N_\text{ey}},\,\frac{L_z}{N_\text{ez}}\right)$ and $2\sqrt{3}\times \max\left(\frac{L_x}{N_\text{ex}},\,\frac{L_y}{N_\text{ey}},\,\frac{L_z}{N_\text{ez}}\right)$, and  $\left\{0.040,\, 12\right\}$ and $\left\{0.001,\,8\right\}$ respectively. Other design parameters are the same as those used in the above sections.

Figure~\ref{fig:3Darchresult} and Fig.~\ref{fig:3Dtowerresult} show the optimized arch and tower designs in different view directions respectively. An isosurface with the physical density value at 0.90 is taken to plot the optimized results. The results in an isoview direction are displayed in Fig.~\ref{fig:3DarchV1} and Fig.~\ref{fig:3DTV1} respectively. It can be noted that $+y-$direction view of the optimized 3D arch result (Fig.~\ref{fig:3Darchy+}) resembles that obtained for the 2D arch structure (Fig.~\ref{fig:archsolution}). For both  designs, the optimizer has succeeded in providing the optimized shape which can be more advantageous for the compliance objective. The material distribution pertaining to different cross-sections for the arch and tower structures are shown in Fig.~\ref{fig:3Darchslice} and Fig.~\ref{fig:3Dtowerslice} respectively. It can be noticed that the material distributions at the cross-sectional areas are close to 1.0  indicating that the optimized results are closed to binary. In addition, the results render clear design interpretation. We can expect relatively more detailed  features in the optimized tower design with much finer mesh which can be efficiently obtained using fully parallelized framework, which forms one of the future works.

\section{Closure}\label{sec:closure}

This paper presents a novel density-based topology optimization approach to optimize continua involving self-weight. The robustness and versatility of the approach are demonstrated by optimizing various  2D and 3D structures subjected to self-weight. The compliance of the structure is minimized using the Method of Moving Asymptotes with a given volume constraint and a conceptualized constraint.

When the effects of self-weight dominate, the optimization problem tends to become unconstrained. We conceptualize a constraint using the maximum permitted mass and intermediate mass of the design domain that implicitly imposes the lower bound on the given volume fraction. With this constraint, the given volume fraction for the problem gets satisfied and remains active at the end of the optimization, i.e., the constrained nature of the problem is retained. With self-weight, the optimized results are significantly different and thus, it should not be neglected in design problems wherein effects of self-weight prevail. The mass density of each element is proposed to interpolate using a smooth Heaviside projection function that offers continuous transition between the phases of elements as topology optimization advances. Using the proper mass density parameters, the non-monotonous behavior of the objective is tuned/controlled.  The mass density parameters are selected \textit{a priori} to the optimization that affect the optimized topologies. The recommended range of values for them are provided based on the numerical examples solved.

The modified SIMP material interpolation method is employed in conjunction with a three-field density representation technique. Parameter $\beta$ is increased using a continuation scheme such that the optimized designs can be steered toward close to 0-1 solutions. The presented approach works well with designs containing non-design domains, which is demonstrated via a house arch design containing a non-design void region. The approach maintains symmetric nature of problem in the optimized structures. The approach is easily extended to 3D problem settings that is demonstrated by solving two three-dimensional problems (a 3D arch structure experiencing only self-weight and a 3D tower structure experiencing both self-weight and external load). It is noted that convergence of the objective is smooth and rapid while obtaining the solutions close to 0-1, i.e. the parasitic effects are subdued. The optimized results provide clear design interpretation. Design problems with self-weight have non-convex nature and therefore, convergence to the global optimum and dependence of the optimized solutions on the starting guesses cannot be ensured. Extending the approach with advanced constraints, e.g., stress constraint, bucking constraint forms future research directions.

	\section*{Acknowledgment}
	The author would like to thank Professor G. K. Ananthasuresh for fruitful discussions, Professor  Krister Svanberg for providing MATLAB codes of the MMA optimizer and acknowledge financial support from the Science \& Engineering research board, Department of Science and Technology, Government of India under the project file number RJF/2020/000023. 
	\bibliography{myreference}
	\bibliographystyle{spbasic} 
\end{document}